\documentclass[11pt,a4paper]{article}
\usepackage{amssymb,a4wide}
\usepackage{graphicx}
\usepackage[small]{caption}
\author{
W. de Boer, C. Sander\\
{\it Institut f\"ur Experimentelle Kernphysik, University of Karlsruhe}\\
Postfach 6980, D-76128 Karlsruhe, Germany
}
\title{Global Electroweak Fits and Gauge Coupling Unification}

\newlength{\dslashwidth}

\newcommand{\bsg}{\ensuremath{b\to X_s\gamma}}

\newcommand{\sinw}{\ensuremath{\sin^2\theta_W}}
\newcommand{\Mgut}{\ensuremath{M_{\mbox{\scriptsize{GUT}}}}}
\newcommand{\agut}{\ensuremath{\alpha_{\mbox{\scriptsize{GUT}}}}}
\newcommand{\tb}{\ensuremath{\tan\beta}}

\begin{document}

\maketitle

\begin{abstract}

%  A global statistical $\chi^2$ analysis of all electroweak data including new data on the anomalous magnetic moment of the muon and the \bsg~ decay rate in both the SM and the MSSM has been performed. The total $\chi^2$ of the MSSM is better than in the SM, mainly because of $a_\mu$, although the total probability is not quite larger due to the larger number of parameters in the MSSM. $A_{FB}^b$ and $A_{LR}$ yield different values of \sinw, which are 3$~\sigma$ apart. Rescaling of the corresponding errors do not change the fit parameters. In addition the fit is performed in the supergravity inspired Constrained MSSM (CMSSM). 

%  In the CMSSM the unification of the gauge couplings is possible, but it turns out that the value of the strong coupling constant and the weak mixing angle have to be larger than the averaged values. But perfect unification can be obtained by \sinw~ from $A_{FB}^b$ and $\alpha_s$ from $R_l$ and $R_\tau$. The latter measurements yield $\alpha_s=0.1216(23)$, while the peak cross section $\sigma_{\mbox{\scriptsize{had}}}^0$ yields $\alpha_s=0.1153(40)$.
%  This lower value is shown to be related to the low number of neutrino generations (2~$\sigma$ below $N_\nu=3$).

The purpose of this paper is threefold: a) Check if the rather poor Standard Model (SM) fit to all electroweak data
can be improved in its minimal supersymmetric extension (MSSM); b) Check what constraints present electroweak data has on the parameter space of the supergravity inspired Constrained MSSM (CMSSM); c) Check if with the present high precision of the gauge coupling constants exact gauge coupling unification is still possible.

It turns out that the total $\chi^2$ in the MSSM is better than in the SM, 
but the total probability is similar due to the larger number of 
parameters in the MSSM. The value of $\tan\beta$ in the CMSSM is 
constrained to be above 6.5, while the value of the gaugino masses at the 
GUT scale has to be above $\sim$220 GeV, which corresponds to a lower limit  
on the lightest neutralino (chargino) of ca. 95 (175) GeV.

The question of gauge unification is more a question of which coupling constants one takes: the forward-backward asymmetry of the b-quarks $A_{FB}^b$ and the left-right asymmetry  $A_{LR}$ yield values of \sinw, which are 3$~\sigma$ apart, while the strong coupling constant $\alpha_s$ from the hadronic peak cross section of the $Z^0$ resonance ($\alpha_s=0.115(4)$) and the ratio of hadronic and leptonic widths ($\alpha_s=0.123(4)$) also differ. It is shown that the larger values of $\alpha_s$ and \sinw~ are consistent with perfect gauge unification. The low $\alpha_s$ value from the hadronic cross section is correlated with the fact that the number of neutrino flavours is 2~$\sigma$ below the known value of 3 in the SM, which is shown to correspond to a 3~$\sigma$ deviation in the hadronic cross section. 
  The error on $\sigma_{\mbox{\scriptsize{had}}}^0$ is dominated by the theoretical uncertainty on the luminosity, which is common to all experiments, since all experiments use the same luminosity MC generator.
  Lowering $\sigma_{\mbox{\scriptsize{had}}}^0$ by 3~$\sigma$ brings $\alpha_s$ up to 0.12 and $N_\nu$ up to 3.
\end{abstract}

\section{Introduction}

  Shortly after the first high statistics data from LEP became available, the first measurements of the electroweak and strong coupling constants showed that in the Standard Model (SM) unification is excluded by many standard deviations, while in the Minimal Supersymmetric Extension of the SM (MSSM) unification is perfect, if one leaves the SUSY mass scale as a free parameter in the fit \cite{amaldi}. Such unification can be graphically displayed as the crossing of lines representing the evolution of the inverse of the coupling constants versus the log of the energy, which are straight lines in first order and the second order contributions to the renormalization group equations (RGE) are  so small that even in second order the evolution of the inverse coupling constants looks like linear lines meeting at the unification point. The fitted SUSY mass scale needed for unification was low compared with the GUT scale, thus agreeing with previous estimates
\cite{ellis}-\cite{giunti}.

 It was also shown that such a fit is by no means trivial, even if from the naive argument, that two lines always meet, so three lines can always brought to a single meeting point with one additional free parameter, the SUSY mass scale. However, since new mass scales effect all three couplings simultaneously, unification is only reached in rare cases. E.g. a fourth family with an arbitrary mass scale changes all slopes by the same amount, thus never leading to unification. In the case of Supersymmetry unification can be reached \cite{amaldi2}.

  Later the analysis became more sophisticated by introducing a mass spectrum for the SUSY masses instead of a common mass \cite{carena}-\cite{langacker2}. Nevertheless, with the more precise calculations and the more precise coupling constants it looked like no perfect supersymmetric unification could be obtained anymore \cite{bagger}. However, this analysis considered only a single value of $\sin^2\theta_W$, which needed values of $\alpha_s(M_Z)>0.13$ for unification. As will be shown, newer values of \sinw~ can lead to perfect unification with smaller values of $\alpha_s$. Unfortunately, the two most precise values of $\sin^2\theta_W$ in the world average of the Particle Data Book are about 3~$\sigma$ apart. Furthermore, the theoretically and experimentally most precise values of the strong coupling constant from the hadronic peak crossection and the ratio $R_l$ of the hadronic and leptonic decay widths of the $Z^0$ resonance are also 2~$\sigma$ apart.  These discrepancies are among the ones, which bring the probablility of the Standard Model (SM) fit to all electroweak precision data to a value around 5\%.

Therefore, the purpose of this paper is threefold:

\begin{itemize}
  \item Compare the global SM fit with a corresponding fit in the minimal supersymmetric SM (MSSM) in order to see if the fit can be improved and especially see if the supersymmetric loop corrections can change the value of the fitted gauge coupling constants. It turns out that the values of the coupling constants do not change in the MSSM, but the two and three sigma deviations discussed above stay also in the MSSM.
  \item Study the constraints of the electroweak precision data on the MSSM parameter space, including the constraints from electroweak symmetry breaking (Constrained MSSM or CMSSM).
  \item Reconsider the gauge unification as funtion of the electroweak mixing angle $\sin^2\theta_W$ and the strong coupling constant $\alpha_s$. It will be shown that the precise value of $\sin^2\theta_W$ from the forward-backward asymmetries of the b-quark together with the precise value of the strong coupling constant from the ratio of the hadronic and leptonic widths of the $Z^0$ boson yield perfect unification, while other precise values do not.
\end{itemize}

The paper is organized as follows: Section 2 summarizes the experimental data used in the fit. In section 3 the global fits in the SM, MSSM and CMSSM are discussed, while section 4 summarizes the constraints on the CMSSM parameter space. Section 5 discusses the gauge unification with the gauge couplings determined from the global fits.

\section{Experimental Data}

  A summary of the most recent electroweak data from colliders can be found in the report of the Electroweak Working Group (EWWG) \cite{ewwgreport}. We included in addition the anomalous magnetic moment of the muon $a_\mu$, which was determined by the E821 collaboration from a measurement of $g-2$ using the polarization in the decays of muons in a muon storage ring. Originally they found $a_\mu$ to be slightly above the SM prediction, the difference being $\Delta a_\mu=a_\mu^{\mbox{\scriptsize{exp}}}-a_\mu^{\mbox{\scriptsize{theo}}}=(430\pm 160)\cdot 10^{-11}$ \cite{e821}. However, the SM expectation $a_\mu^{\mbox{\scriptsize{theo}}}$ was recently increased, which reduces the discrepancy in $\Delta a_\mu$ to ($176\pm 170)\cdot 10^{-11}$ \cite{narison}. The reasons for the increase in the SM expectation were several: a sign error in the hadronic light-by-light scattering contribution was found \cite{lightbylight}, secondly new data on the low energy hadronic cross section became available \cite{bes} and thirdly in the value above the data on hadronic $\tau$-decays were used to calculate the vacuum polarization correction to the fine structure constant using CVC (conserved vector current). Recently new data on the anomalous magnetic moment were published \cite{e821new}. The new value of $\Delta a_\mu$ corresponds to a $1.6$ to $3.0~\sigma$ deviation from the SM, depending on which SM prediction is used \cite{DH}. If the hadronic $\tau$-decays are used one obtains a higher SM prediction, which results in a lower deviation ($1.6~\sigma$). On the other hand, if the new data on the low energy cross section for the $e^+e^-$-annihilation into hadrons is used one obtains $\Delta a_\mu=(339\pm 112)\cdot 10^{-11}$ (corresponding to a $3.0~\sigma$ deviation). We will repeat the fits with the low (1.6~$\sigma$) and high (3.0~$\sigma$) value of $\Delta a_\mu$ denoted by $\Delta a_\mu^{\mbox{\scriptsize{small}}}$ and $\Delta a_\mu^{\mbox{\scriptsize{large}}}$.
%Only with this $3.0~\sigma$ deviation one can obtain 95\% C.L. upper limits on the masses, which will be indicated in this paper, although one should keep in mind, that these upper limits disappear, when the hadronic $\tau$-decays are used.

  SUSY contributions are also expected to affect the \bsg~ rate, for which the most recent world average is: $Br(\bsg)=(3.43\pm 0.35)\cdot 10^{-4}$. This value is dominated by the recently published results from BaBar ($(3.88\pm 0.36_{\mbox{\scriptsize{stat}}}\pm 0.37_{\mbox{\scriptsize{syst}}}\pm 0.36_{\mbox{\scriptsize{mod}}})\cdot 10^{-4}$) \cite{barbar} and CLEO ($(3.21\pm 0.43_{\mbox{\scriptsize{stat}}}\pm 0.27_{\mbox{\scriptsize{syst}}}\pm 0.14_{\mbox{\scriptsize{mod}}})\cdot 10^{-4}$) \cite{cleo}, which are consistent with the results from ALEPH \cite{aleph} and BELLE \cite{belle}. The latter have considerably larger errors.

  The world average is slightly below, but consistent with a recent SM prediction by Gambino and Misiak of $(3.73\pm 0.30)\cdot 10^{-4}$ \cite{bsgtheo}. This value is somewhat higher than previous predictions, since it uses the running mass for the charm quark in the loops, while keeping the pole mass for the bottom quark in the external lines. This gives an additional uncertainty, but the authors found a reduced scale dependence. In our present analysis we conservatively keep a theoretical error of $\pm 0.40\cdot 10^{-4}$, but use $m_c(\mu)/m_b=0.22$. This is not critical, since with the present large experimental error \bsg~ hardly constrains the fit.

\section{Comparison of Fits to Electroweak Precision Data in the SM, MSSM and CMSSM}

  A few years ago a complete electroweak fit program including all possible supersymmetric corrections in the Minimal Supersymmetric Model (MSSM) was developed, mainly to investigate the so-called $R_b$ deviation of the Standard Model (SM) \cite{mssmfitter}. At present $R_b$ shows no significant deviation from the SM, but the present total $\chi^2$ of all electroweak data is not excellent \cite{ewwgreport}, especially if the new measurements of the anomalous magnetic moment of the muon $a_\mu$ \cite{e821} and \bsg~ \cite{barbar},\cite{cleo} are included. None of these latter measurements show by itself a significant deviation from the SM, but since they all point to supersymmetric contributions, it is interesting to compare a global fit of all data in the SM, MSSM and CMSSM.

  The fits to the electroweak precision data are performed in three different models:
  \begin{itemize}
    \item Standard Model (SM) with 5 parameters: $\alpha_s(M_Z)$, $M_Z$, $m_t$, $m_h$ and $\Delta\alpha^{(5)}_{\mbox{\scriptsize{had}}}$
    \item Minimal Supersymmetric Model (MSSM): In the most general case all sfermion masses can be chosen independently, because they are not constrained by GUT relations. For simplification we assume a common slepton and common squark mass scale with the exception of the left and right handed stop mass. In the third generation sfermion sector mass splitting due to Yukawa couplings is taken into account. The chargino and neutralino matrices have as free parameters $\mu$ and $M_2$, while $M_1$ was taken to be ${5 \over 3}{\sinw\over\cos^2\theta_W}M_2$, as expected from Renormalization Group Equations (RGEs). For details see Ref. \cite{mssmfitter}.
    \item Constrained Minimal Supersymmetric Model (CMSSM): In this model Supersymmetry is broken by gravity mediation (mSUGRA). This lowers the number of free parameters. \Mgut~ is determined by gauge unification ($\alpha_1=\alpha_2=\alpha_3=\agut$). Because of the discrepancies in the coupling constants from electroweak data, we do not insist on exact gauge unification at \Mgut~ as a constraint in the electroweak fits, but consider the question of gauge unification in the last section.  The sfermion masses are unified at the GUT scale \Mgut~ to $m_0$, just as the gaugino masses are unified to $m_{1/2}$, while $\vert\mu\vert$ is determined by electroweak symmetry breaking (EWSB). Also the trilinear couplings are unified to $A_0$ at the GUT scale. The low energy values are determined by the RGEs \cite{gut}.
  \end{itemize}

%  All electroweak variables were calculated in the SM using ZFITTER 6.11 \cite{zfitter} and in the MSSM using MSSMFITTER \cite{mssmfitter}. The fact that the SM fit prefer a Higgs mass slightly below the experimental limit is not an issue either, since requiring the Higgs mass to be above the limit from direct searches ($m_h>114.6$ GeV) only causes a minor shift in the top mass in the fit (see Tab. \ref{smtab}).

  All electroweak variables were calculated in the SM using ZFITTER 6.11 \cite{zfitter} and in the MSSM using MSSMFITTER \cite{mssmfitter}. Extensive references for the supersymmetric contributions to $\Delta a_\mu$ and \bsg~ can be found in \cite{bsgamu}. The fitted SM parameters are shown in Table \ref{smtab}. From the left column one observes that the preferred Higgs mass is $88^{+54}_{-35}$ GeV. This value is below the experimental limit of 114.6 GeV. If one adds this Higgs limit as a constraint to the $\chi^2$ function\footnote{The Higgs limit was implemented by adding a term $(m_h-116\mbox{~GeV})/(0.986\mbox{~GeV})^2$ to the $\chi^2$ distribution, if the calculated Higgs mass is below 116 GeV.}, the $\chi^2$ becomes hardly worse, since the fit  increases
   the top mass slightly in that case (see second column of Tab. \ref{smtab}).

\begin{table} [tbp]
 \begin{center}
  \begin{tabular}{|c|c|c|c|}
   \hline
   Parameter   & SM               & SM + Higgs    & SM + Higgs + rescaled \\
   \hline
   \hline
   $M_Z$ [GeV] & 91.1874(20)      & 91.1874(21)   & 91.1875(21)\\
   $m_t$ [GeV]
               & 175.5(4.1)       & 177.4(3.0)    & 177.7(3.1)\\
   $m_h$ [GeV] & $88^{+54}_{-35}$ & 114.7         & 114.6\\
   $\alpha_s(M_Z)$
               & 0.1183(26)       & 0.1185(26)    & 0.1182(27)\\
   $\Delta\alpha^{(5)}_{\mbox{\scriptsize{had}}}$
               & 0.02770(31)      & 0.02761(31)   & 0.02755(33)\\
   \hline
   \hline
   $\sin^2\theta ~(\overline{\mbox{\scriptsize{MS}}})$
               & 0.23136(16)      & 0.23143(10)   & 0.23139(12)\\
   $M_W$ [GeV] & 80.399(18)       & 80.397(17)    & 80.400(19)\\
   \hline
   \hline
   $\chi^2$/d.o.f. ($\Delta a_\mu^{\mbox{\scriptsize{large}}}$)
               & 26.9/16          & 27.2/16       & 21.0/16\\
   Probability ($\Delta a_\mu^{\mbox{\scriptsize{large}}}$) 
               & 4.7\%            & 3.9\%         & 18.0\% \\
   Probability ($\Delta a_\mu^{\mbox{\scriptsize{small}}}$)
               & 22.3\%           & 20.9\%         & 59.6\% \\
   \hline
  \end{tabular}
  \caption[]{The five SM fit parameters: the $Z^0$ mass $M_Z$, the top mass $m_t$, the Higgs mass $m_h$, the strong coupling constant at $M_Z$  $\alpha_s(M_Z)$ and the hadronic contribution to the electromagnetic coupling constant $\Delta\alpha^5_{\mbox{\scriptsize{had}}}$. The derived quantities $\sin^2\theta ~(\overline{\mbox{\scriptsize{MS}}})$ and the mass of the charged weak boson $M_W$ are listed in addition. The lowest rows indicate the $\chi^2$/d.o.f. and the corresponding probability. The probability in the last line is obtained if hadronic $\tau$-decays are used to get a small $\Delta a_\mu$. The first column uses the input data discussed in the text, which yields as most probable Higgs mass $m_h$=88 GeV; the second column requires the Higgs mass to be above the experimental limit of 114.6 GeV at 95\% C.L.; in the third column the errors of $A_{FB}^b$ and $A_{LR}$ are rescaled according to the Particle Data Group prescription in order to resolve the 3$~\sigma$ discrepany. The fit parameters are hardly changed by this rescaling.}\label{smtab}
 \end{center}
\end{table}

\begin{figure} [tbp]
  \begin{center}
    \includegraphics[width=0.49\textwidth]{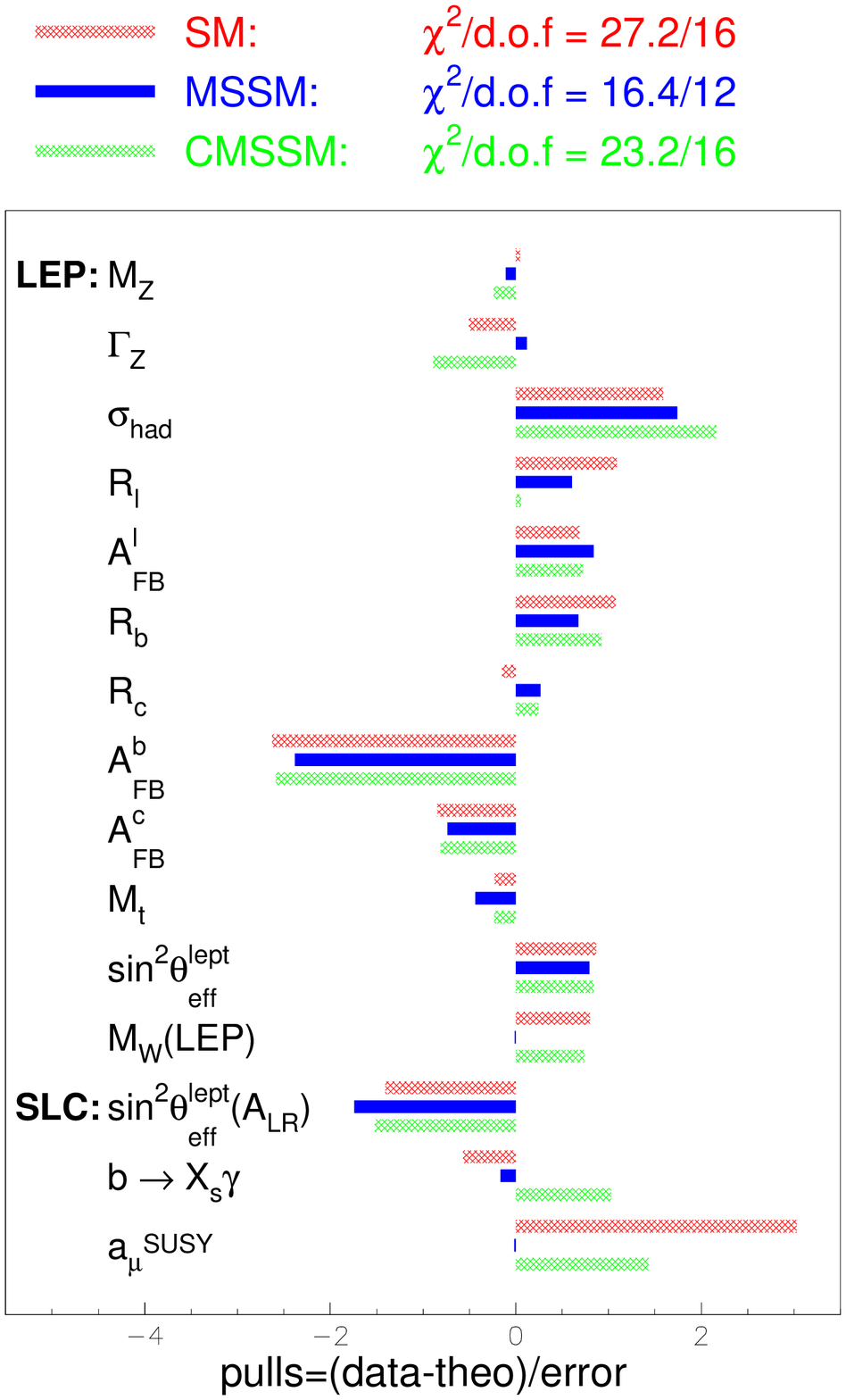}
    \includegraphics[width=0.49\textwidth]{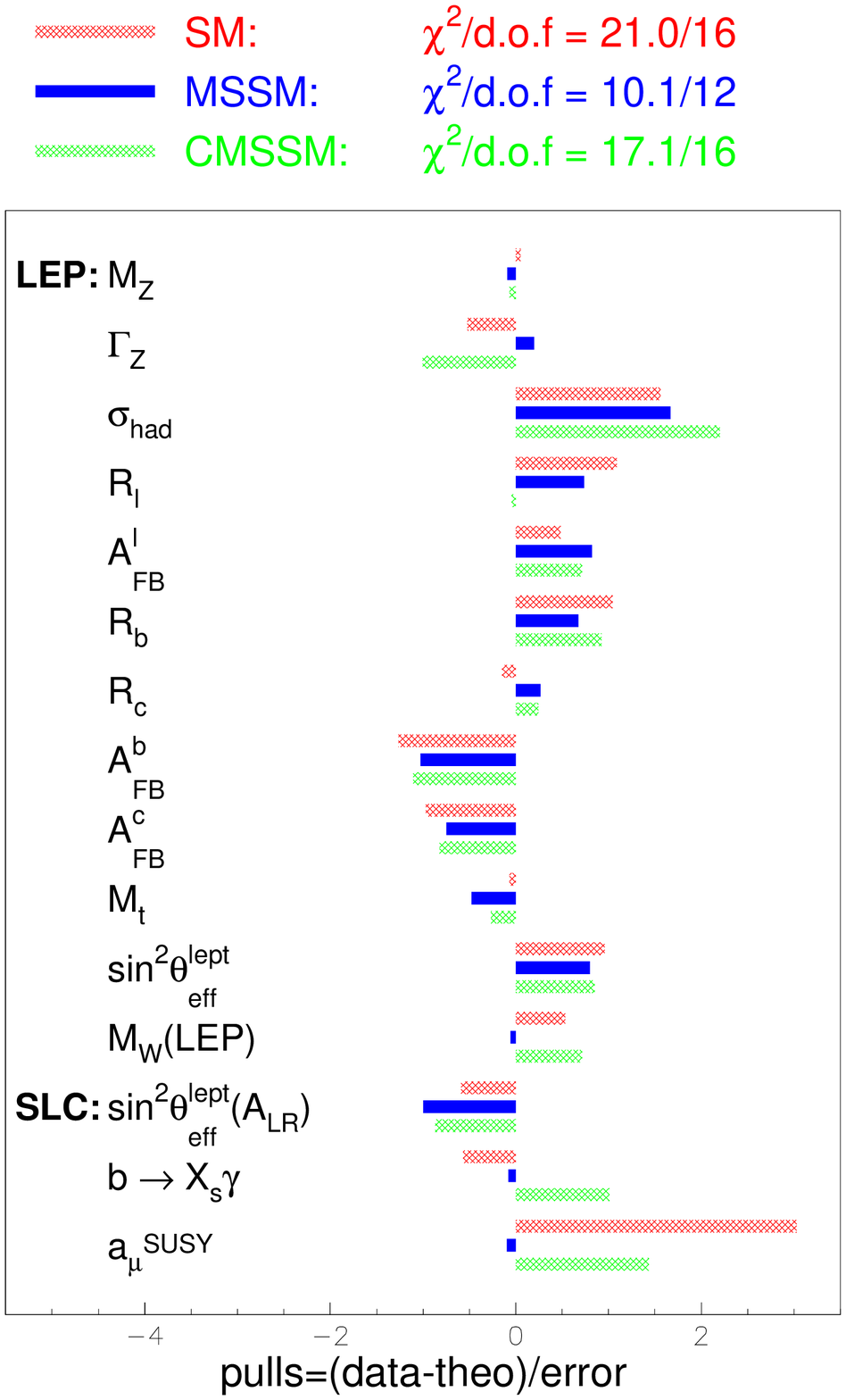}
    \caption[]{The pulls of the electroweak data for the SM, MSSM and CMSSM. On the left hand side the full errors are taken into account. On the right hand side the errors for $A_{FB}^b$ and $A_{LR}$ are rescaled as described in the text. The total $\chi^2$ is in the MSSM better than in the SM, due to $a_\mu$, but the probabilities are similar (see Tab. \ref{smtab} and \ref{mssmtab})} \label{pulls}
  \end{center}
\end{figure}

The bottom row of Tab. \ref{smtab} shows that the probability of the SM varies between 5\% and 22\%, depending on the value of $\Delta a_\mu$. The deviations between various observables and the theoretical predictions from the SM and the MSSM models are shown in Fig. \ref{pulls} in units of standard deviations (``pulls''). The variables with the largest pulls are: the forward-backward asymmetry $A_{FB}^b$, the left-right asymmetry $A_{LR}$, the anomalous magnetic moment of the muon $\Delta a_\mu$ and the hadronic peak cross section $\sigma_{\mbox{\scriptsize{had}}}^0$ of the $Z_0$ peak. 

%The direct measurements of $M_W$ at LEP II and $p\bar{p}$ colliders yield a larger value than the SM prediction from LEP I. If SUSY corrections are included, the latter prediction increases, as shown in Fig. \ref{mw}. The SM value of $M_W$ is a function of $M_Z$, \sinw, $m_t$ and $m_h$ without a constraint from the direct measurement. With the latter constraint included one gets a somewhat higher value (see Tab. \ref{smtab}). It should be mentioned that the $W$ mass from LEP II is still preliminary.
%\begin{figure}
%  \begin{center}
%    \includegraphics[width=0.5\textwidth]{mw_msf.eps}
%    \caption[]{$W$ mass versus sparticle masses, assuming all sparticles have the same mass. The horizontal bands represent the SM prediction from LEP I data and the direct measurement from LEP II and $p\bar{p}$ colliders. The curved band is the MSSM prediction for the case that all sparticles have a given mass $m_{\mbox{\scriptsize{SUSY}}}$. Its width is determined by the uncertainty from the top mass.} \label{mw}
%  \end{center}
%\end{figure}

\begin{figure} [tbp]
  \begin{center}
    \includegraphics[width=0.6\textwidth]{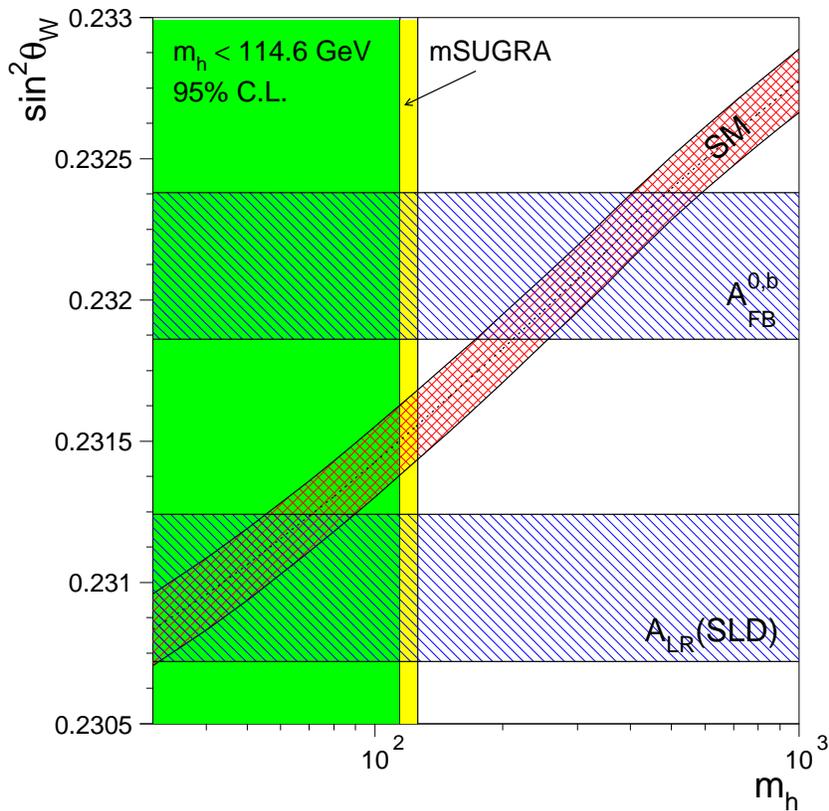}
    \caption[]{The dependence of the weak mixing angle on the Higgs mass in the SM. The mixing angle measured at SLAC (lower horizontal bar) prefers a low Higgs mass, which is excluded by direct searches at LEP. The uncertanty of the SM prediction is given by the uncertanty of the top mass. The vertical line around 120 GeV is the expectation from the CMSSM.} \label{sw2mh}
  \end{center}
\end{figure}

The measurements of the forward-backward asymmetry $A_{FB}^b$ for $b$-quarks and the left-right asymmetry $A_{LR}$, as measured with the polarized electron beam at SLAC, can be both translated into measurements of the electroweak mixing angle, which than turns out to be 3$~\sigma$ apart \cite{ewwgreport}. The  weak mixing angle is strongly correlated with the Higgs mass, as shown in Fig. \ref{sw2mh}. The value of $\sinw=0.23098(26)$ measured at SLAC prefers a Higgs mass around 40 GeV, which is lower than the present Higgs limit of 114.6 GeV \cite{higgslimit}. The value of $\sinw=0.23226(31)$ measured at LEP prefers a much larger Higgs mass. However, the errors are large, so requiring $m_h>114.6$ GeV does not worsen the fit significantly, as can be seen from Table \ref{smtab}. So the fact that $A_{LR}$ corresponds to a too low Higgs value is not a strong argument to disfavour it. Therefore, we followed the procedure from the Particle Data Group to rescale the errors of these variables in such a way that their $\chi^2$ contributions are about one \cite{pdg}. This hardly influences any of the fit parameters, as shown in Fig. \ref{pulls} and Tables \ref{smtab} and \ref{mssmtab}, but increases the probability from 5\% (17\%) to 18\% (61\%) in the SM (MSSM). Since the fitted parameters are hardly affected by the rescaling, this is not a critical issue.

\begin{table} [tbp]
 \begin{center}
  \begin{tabular}{|c|c|c|c|c|c|c|}
   \hline
   & \multicolumn{3}{|c|}{full errors} & \multicolumn{3}{|c|}{rescaled errors}\\
   \hline
          & $\tan\beta=20$& $\tan\beta=35$& $\tan\beta=50$
          & $\tan\beta=20$& $\tan\beta=35$& $\tan\beta=50$\\
   \hline
   $m_t$ [GeV]                       & 175.4   & 176.5   & 176.9   & 175.6   & 176.7   & 177.9   \\
   $\alpha_s$                        & 0.1187  & 0.1189  & 0.1186  & 0.1186  & 0.1184  & 0.1185  \\
   $\mu$ [GeV]                       & 129.4   & 132.0   & 132.0   & 129.3   & 131.8   & 132.8   \\
   $m_{\tilde t_1}$ [GeV]            & 561     & 604     & 635     & 573     & 608     & 636     \\
   $m_{\tilde t_2}$ [GeV]            & 557     & 602     & 635     & 569     & 606     & 635     \\
   $m_{\tilde b_1}$ [GeV]            & 540     & 588     & 627     & 552     & 592     & 627     \\
   $m_{\tilde b_2}$ [GeV]            & 458     & 504     & 555     & 464     & 516     & 555     \\
   $m_{\tilde q}$ [GeV]              & 540     & 588     & 627     & 552     & 592     & 627     \\
   $m_{\tilde l}$ [GeV]              & 540     & 586     & 705     & 384     & 576     & 705     \\
   $ m_{\tilde{\chi}_1^0}$ [GeV]     & 68.7    & 68.3    & 68.1    & 68.7    & 68.2    & 69.0    \\
   $m_h$ [GeV]                       & 116.5   & 118.7   & 119.9   & 117.0   & 118.9   & 120.4   \\
   \hline
   \hline
   $M_W$ [GeV]                       & 80.425  & 80.426  & 80.427  & 80.426  & 80.428  & 80.433  \\
   $\sin^2\theta^{\mbox{\scriptsize{lept}}}_{\mbox{\scriptsize{eff}}}$
                                     & 0.23142 & 0.23145 & 0.23145 & 0.23141 & 0.23144 & 0.23142 \\
   $Br(\bsg)\times 10^{4}$           & 3.49    & 3.52    & 3.40    & 3.46    & 3.50    & 3.47    \\
   $\Delta a_\mu\times 10^{11}$      & 347     & 341     & 365     & 358     & 350     & 365     \\
   \hline
   \hline
   ${\chi^2}$/d.o.f.                 & 15.8/12 & 16.4/12 & 16.9/12 & 9.5/12  & 10.1/12 & 10.5/12 \\
   Probability                       & 19.9\%  & 17.2\%  & 15.5\%  & 66.3\%  & 60.6\%  & 56.9\%  \\
   \hline
  \end{tabular}
  \caption[]{The best fit parameters for different large $\tan\beta$ scenarios are given. The masses are independent, i.e. they were chosen at low energies independent of possible GUT relations. All sparticle masses, which had no influence on the fit, like gluino and pseudo scalar Higgs mass, were set to a large value (1 TeV). $M_2$ was set to 170 GeV and $\mu$ was chosen that way to get a lightest chargino mass of 103.5 GeV which is the present lower limit from direct searches at LEP. All squark masses were chosen to be equal, except for the stop masses. Also all slepton masses were chosen to be equal. In the last 3 columns the fit was repeated after rescaling the errors of $A_{FB}^b$ and $A_{LR}$ (see text).
%Note the increase in $M_W$ compared with the one in Tab. \ref{smtab}.
} \label{mssmtab}
 \end{center}
\end{table}

\begin{figure} [tbp]
  \begin{center}
    \includegraphics[width=0.5\textwidth]{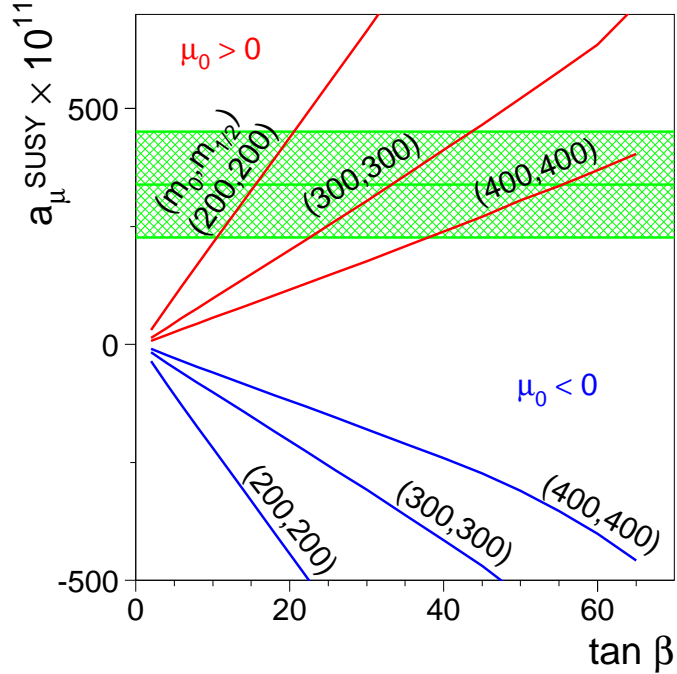}
    \caption[]{Dependence of the anomalous magnetic moment of the muon $a_\mu$ on $\tan\beta$ for different supersymmetric sparticle masses, parameterized by the common GUT scale masses for the spin 0 and spin 1/2 sparticles, called $m_0$ and $m_{1/2}$ respectively. The horizontal band represent the experimental measurement $\Delta a_\mu=(339\pm 112)\cdot 10^{-11}$} \label{amutanb}
  \end{center}
\end{figure}

\begin{figure} [tbp]
  \begin{center}
    \includegraphics[width=0.8\textwidth]{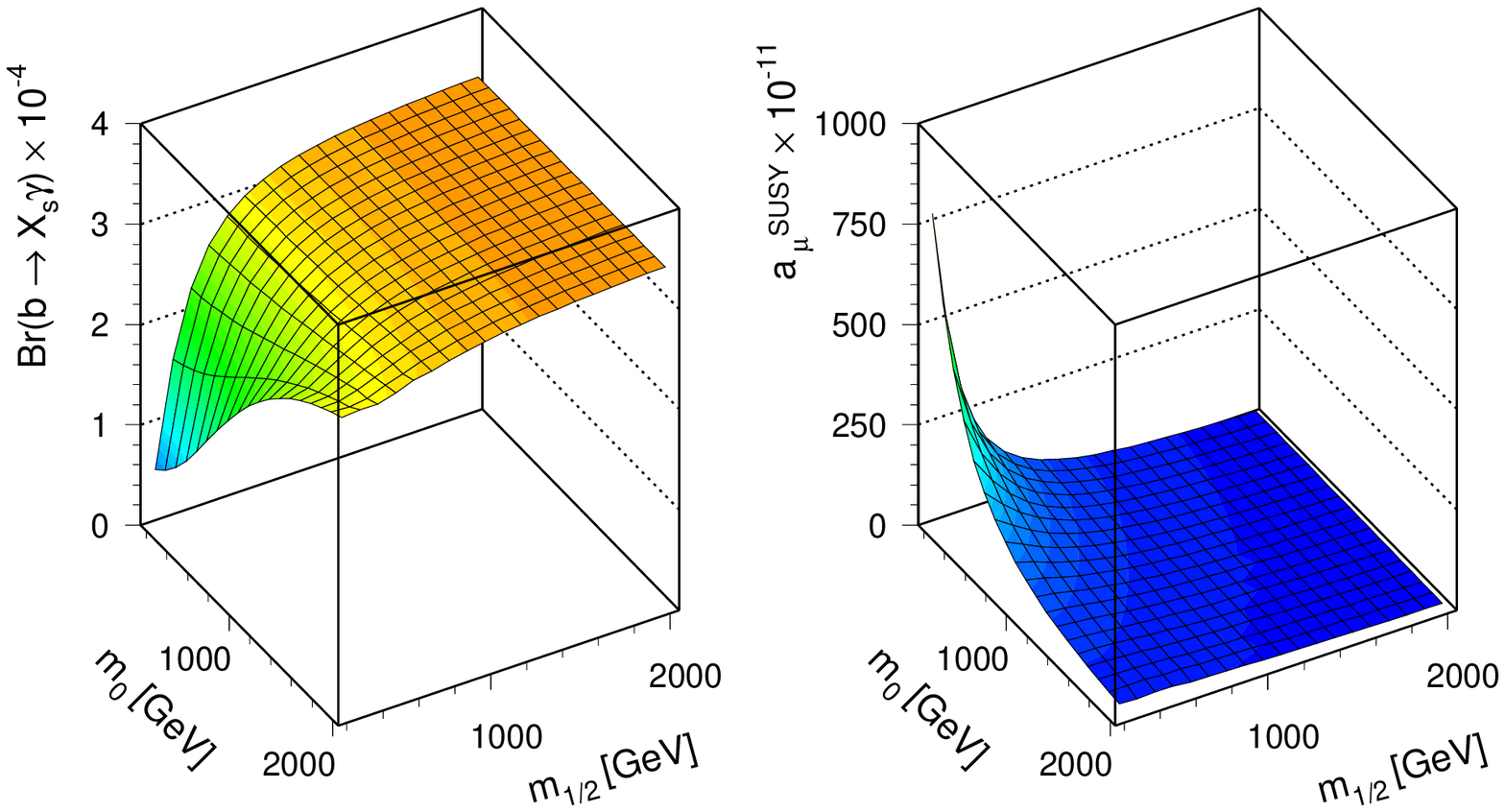}
    \hspace{0.05\textwidth}
    \includegraphics[width=0.3\textwidth]{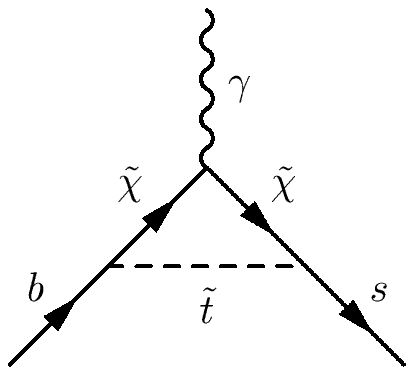}
    \hspace{0.1\textwidth}
    \includegraphics[width=0.3\textwidth]{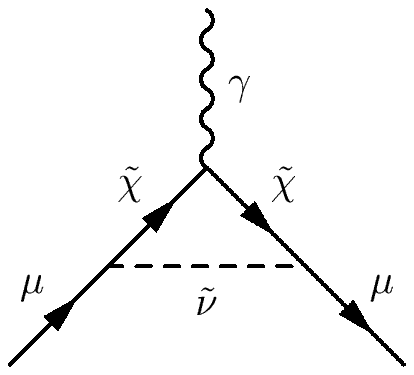}
    \caption[]{The values of \bsg~ and $\Delta a_\mu^{\mbox{\scriptsize{SUSY}}}$ in the $m_0$-$m_{1/2}$-plane for positive $\mu$ and $\tan\beta=35$ to be compared with the experimental data $Br(\bsg)=(3.43\pm 0.36)\cdot 10^{-4}$ and $\Delta a_\mu=(339\pm 112)\cdot 10^{-11}$. One observes that a value of \bsg~ below the SM prediction is correlated with a value of $\Delta a_\mu^{\mbox{\scriptsize{SUSY}}}$ above the SM prediction. The two observables are correlated by the chargino mass, shown in the loops at the bottom.} \label{bsg_amu}
  \end{center}
\end{figure}

  The $\chi^2$/d.o.f. in the MSSM is better than in the SM (16.4/12 for MSSM versus 27.2/16 for SM), mainly because of $a_\mu$ and somewhat because of \bsg~(see Fig. \ref{pulls}), but the probabilitiy is still not good due to the larger number of parameters in the MSSM (17\% for MSSM versus 5\% for SM). The reason for the improved $\chi^2$ are the supersymmetric corrections, which can be large for $a_\mu$, especially for large $\tan\beta$ \cite{cm}, as shown in Fig. \ref{amutanb}. Note the preferred positive sign of $\mu$ and the relatively large value of $\tan\beta$ needed to be consistent with the experimental value of $\Delta a_\mu$. It is interesting to note that the positive sign of $\mu$, as preferred by $a_\mu$, yields indeed a value of \bsg~ slightly below the SM value. This correlation is due to similar loop diagrams for $a_\mu$ and \bsg, both containing charginos as shown in Fig. \ref{bsg_amu}. In addition squark - and slepton masses play a role. These were fitted in addition to the SM parameters using MSSMFITTER. The results are shown in Tab. \ref{mssmtab}. The MSSM fits are not very sensitive to $\tan\beta$, if it is large. As will be shown in the next section, $\tb>6.5$ is required by the combined electroweak data.

  The pulls from a global fit to the electroweak precision data in the CMSSM including  electroweak symmetry breaking, third generation fermion masses and the LSP constraint are shown in Fig. \ref{pulls} in comparison with the SM and MSSM fits. The $\chi^2$ is in the CMSSM larger, but this causes almost no decrease in probability, since the lower number of parameters increases the degrees of freedom.
%(MSSM: 17.0\% $\to$ CMSSM: 6.0\%, 14% without unification).

\clearpage

\section{Constraints on the parameter space in the CMSSM}

\begin{figure} [tbp]
  \begin{center}
    \includegraphics[width=0.6\textwidth]{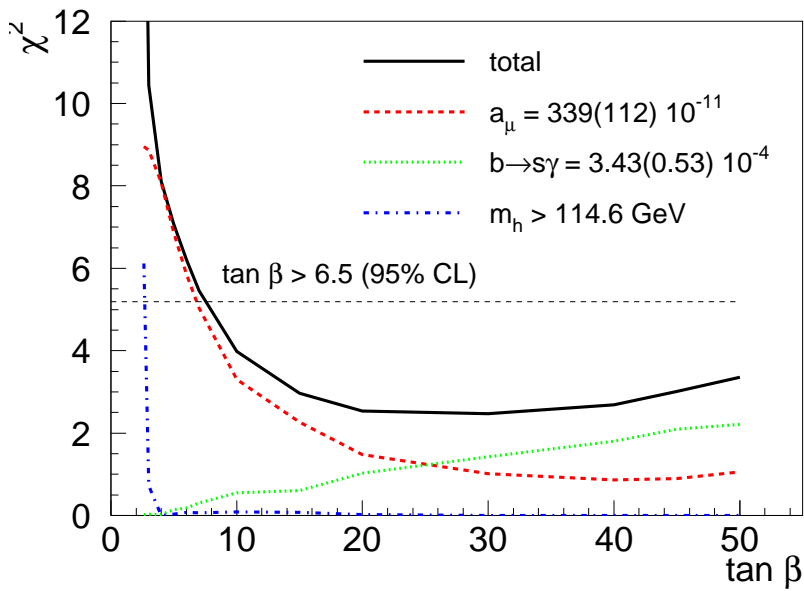}
    \caption[]{From constraits on the anomalous magnetic moment of the muon, the branching ratio $Br(\bsg)$ and the light Higgs mass one can get a lower limit of $\tan\beta$. With the shown constraints we get $\tan\beta>6.5$.} \label{tanblimit}
  \end{center}
\end{figure}

In this section we consider first the constraints on \tb~, from the electroweak data including the 95\% C.L. lower limit on the Higgs mass of 114.6 GeV \cite{higgslimit}. The latter limit is the limit in the SM. However, this is also valid in the CMSSM, since the constraint of EWSB requires a high value of $\mu$, which in turn gives a high mass to the heavier Higgs particles, so the lightest Higgs decouples and obtains the couplings from the SM Higgs. From contraints on $a_\mu$, \bsg~ and the Higgs limit one gets a lower limit on \tb~ of 6.5 at the 95\% C.L., as shown in Fig. \ref{tanblimit}.

\begin{figure} [tbp]
  \begin{center}
    \includegraphics[width=0.4\textwidth]{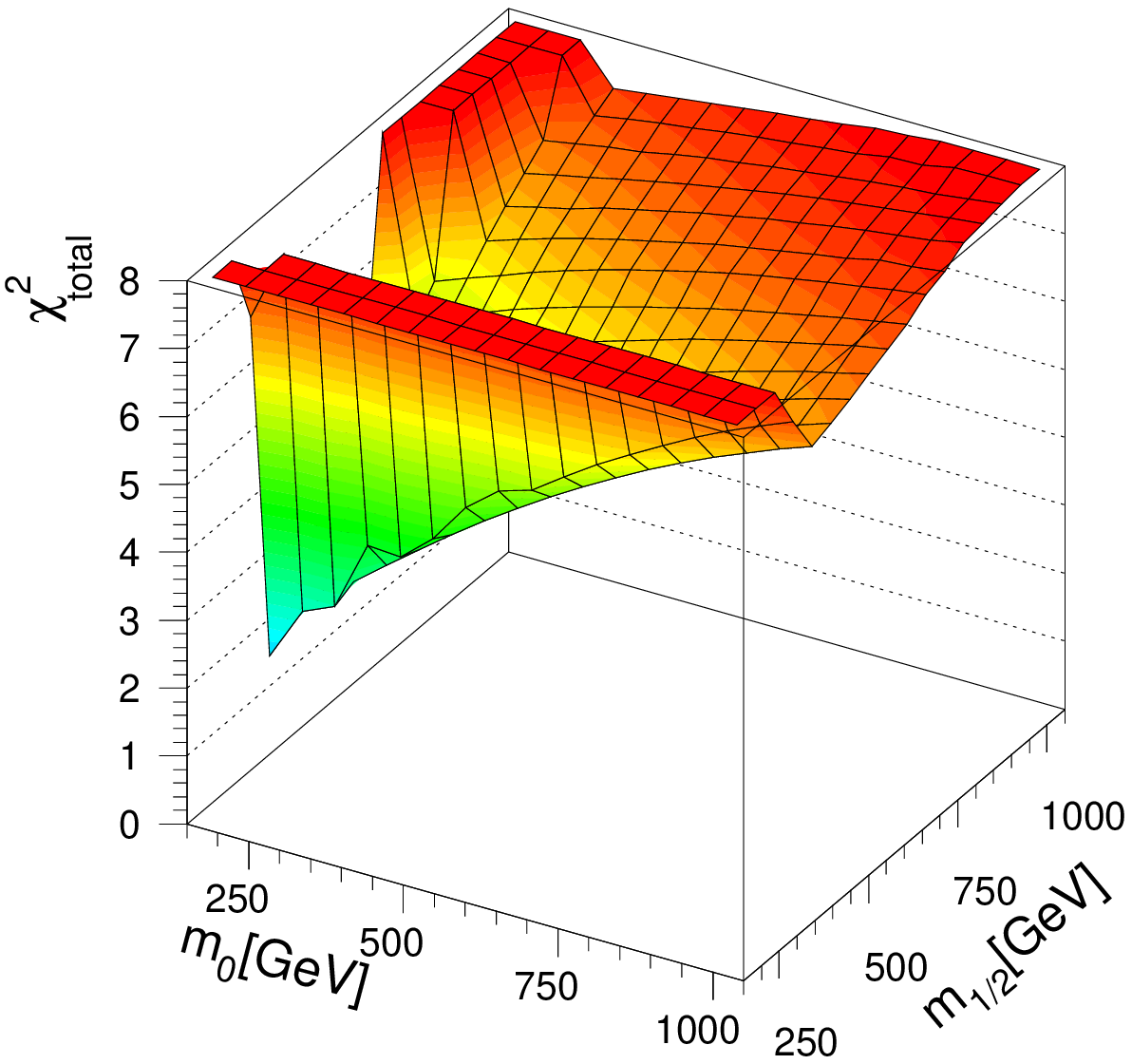}
    \includegraphics[width=0.4\textwidth]{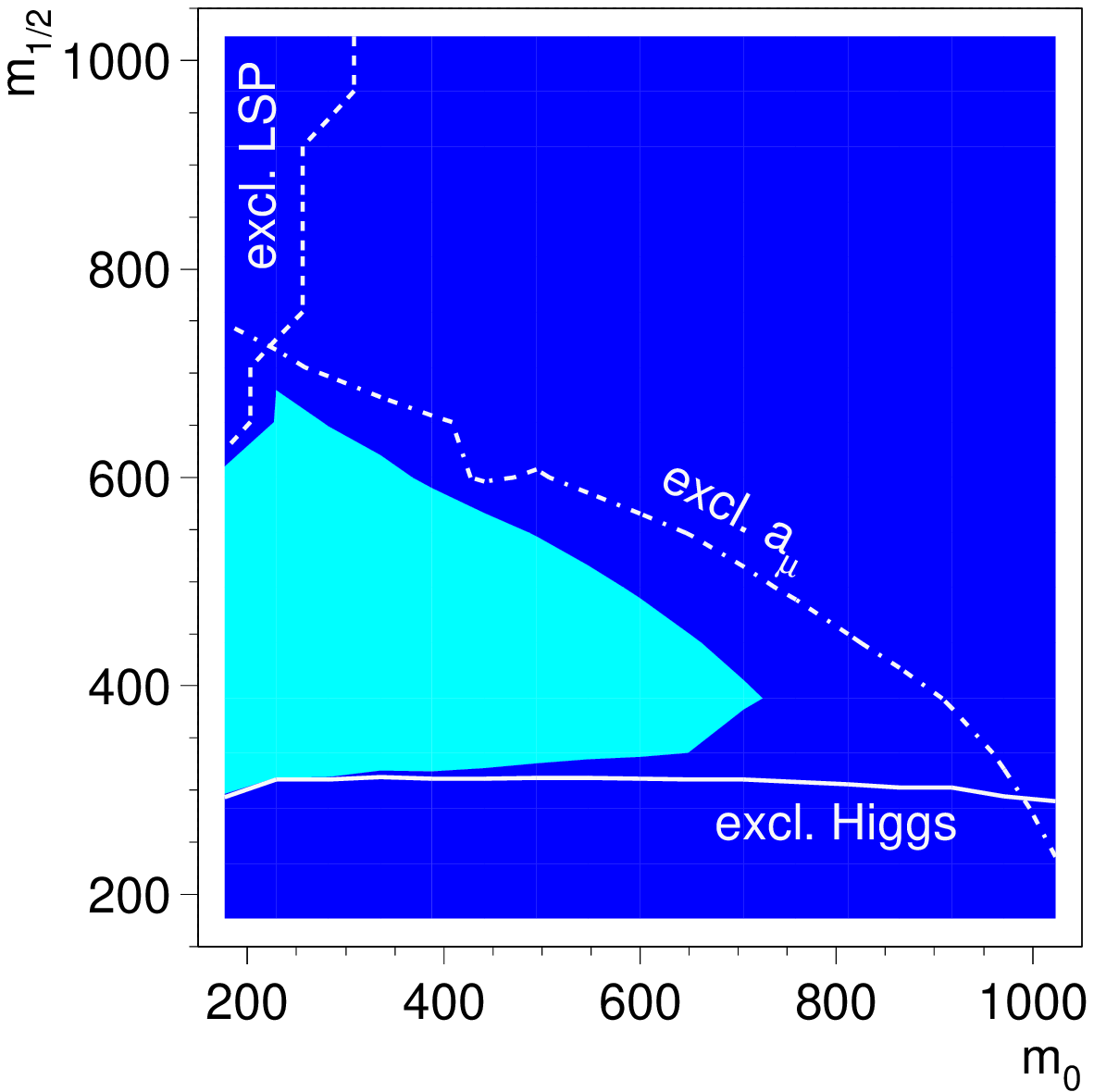}
    \includegraphics[width=0.4\textwidth]{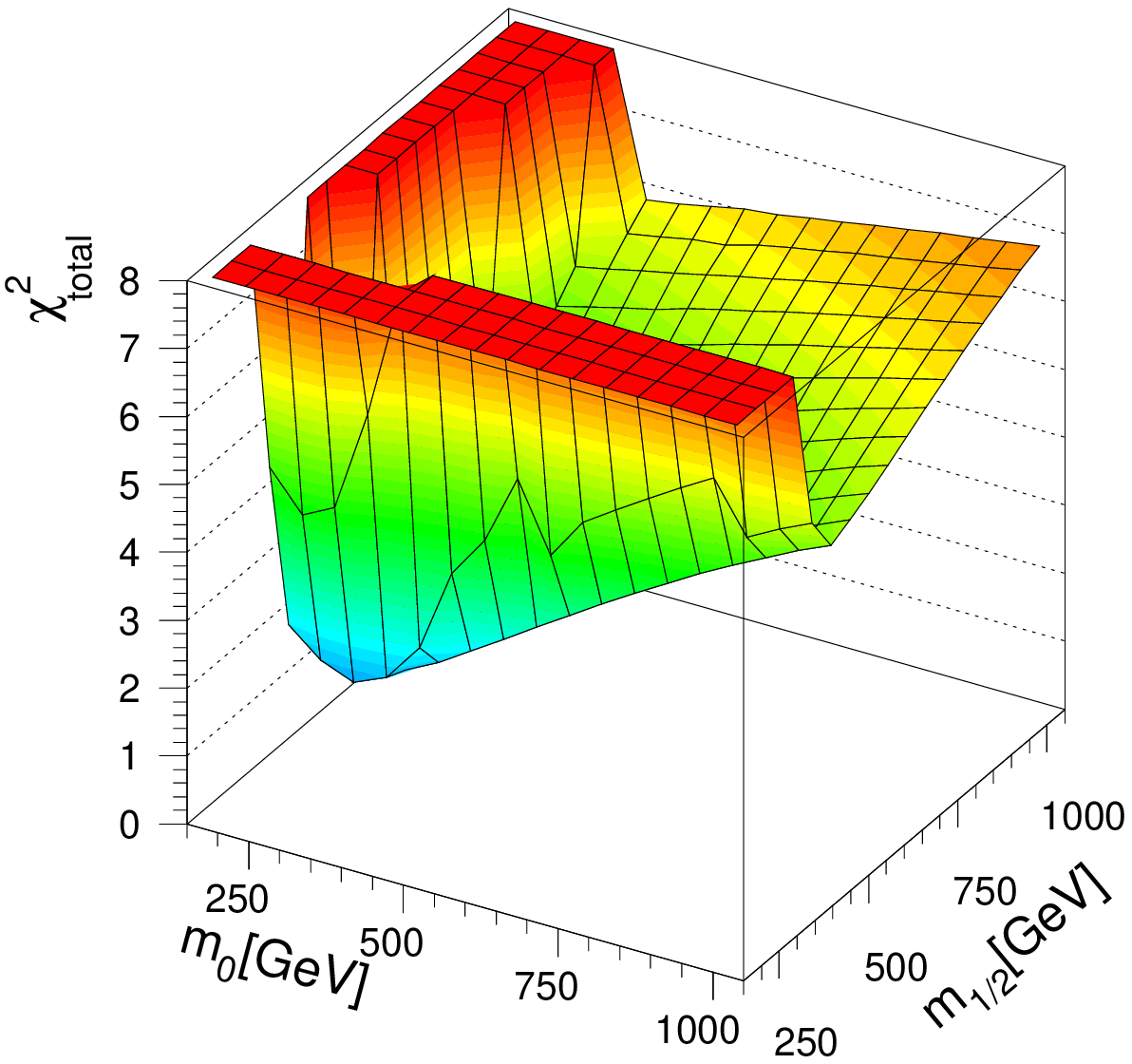}
    \includegraphics[width=0.4\textwidth]{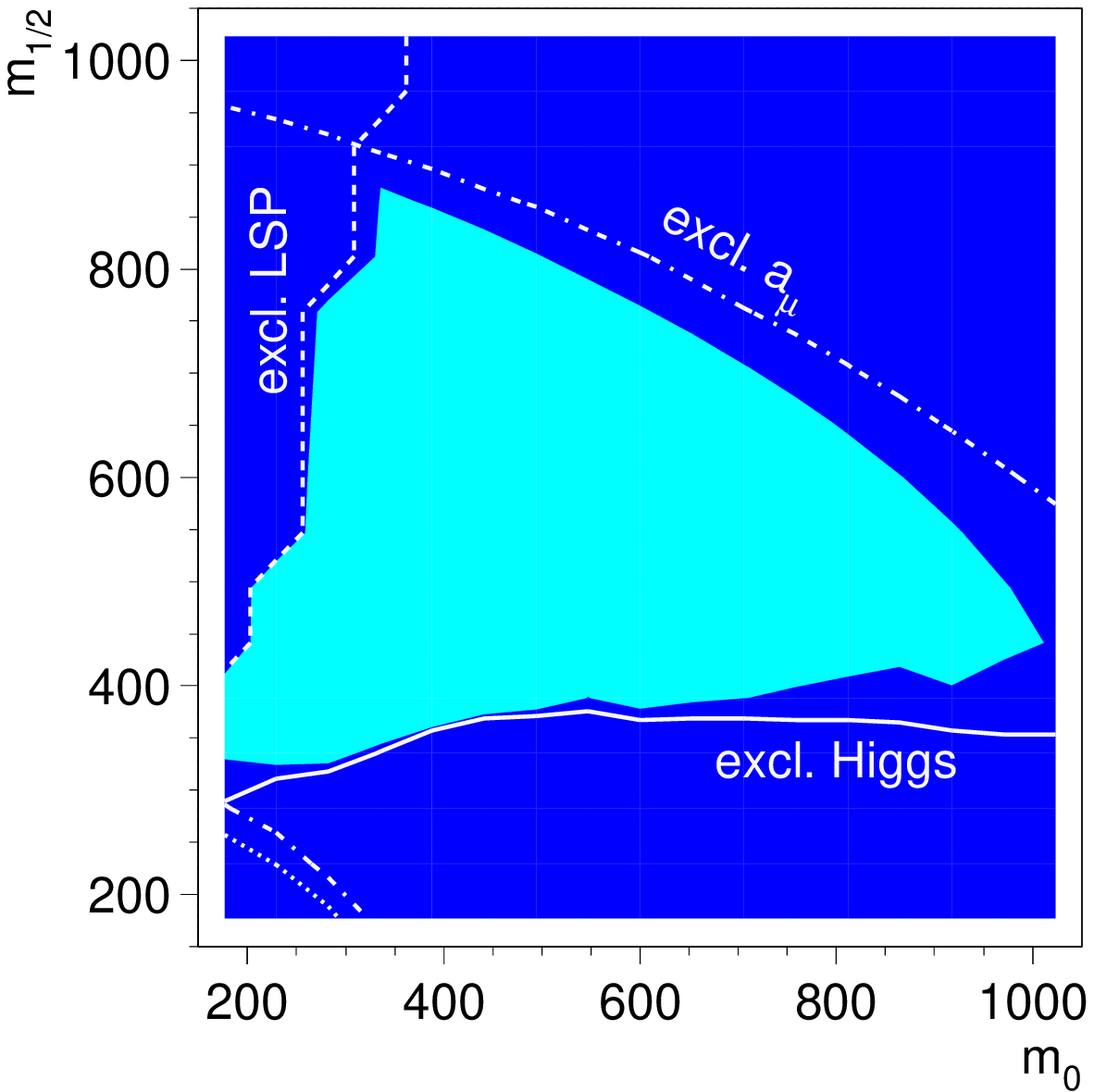}
    \includegraphics[width=0.4\textwidth]{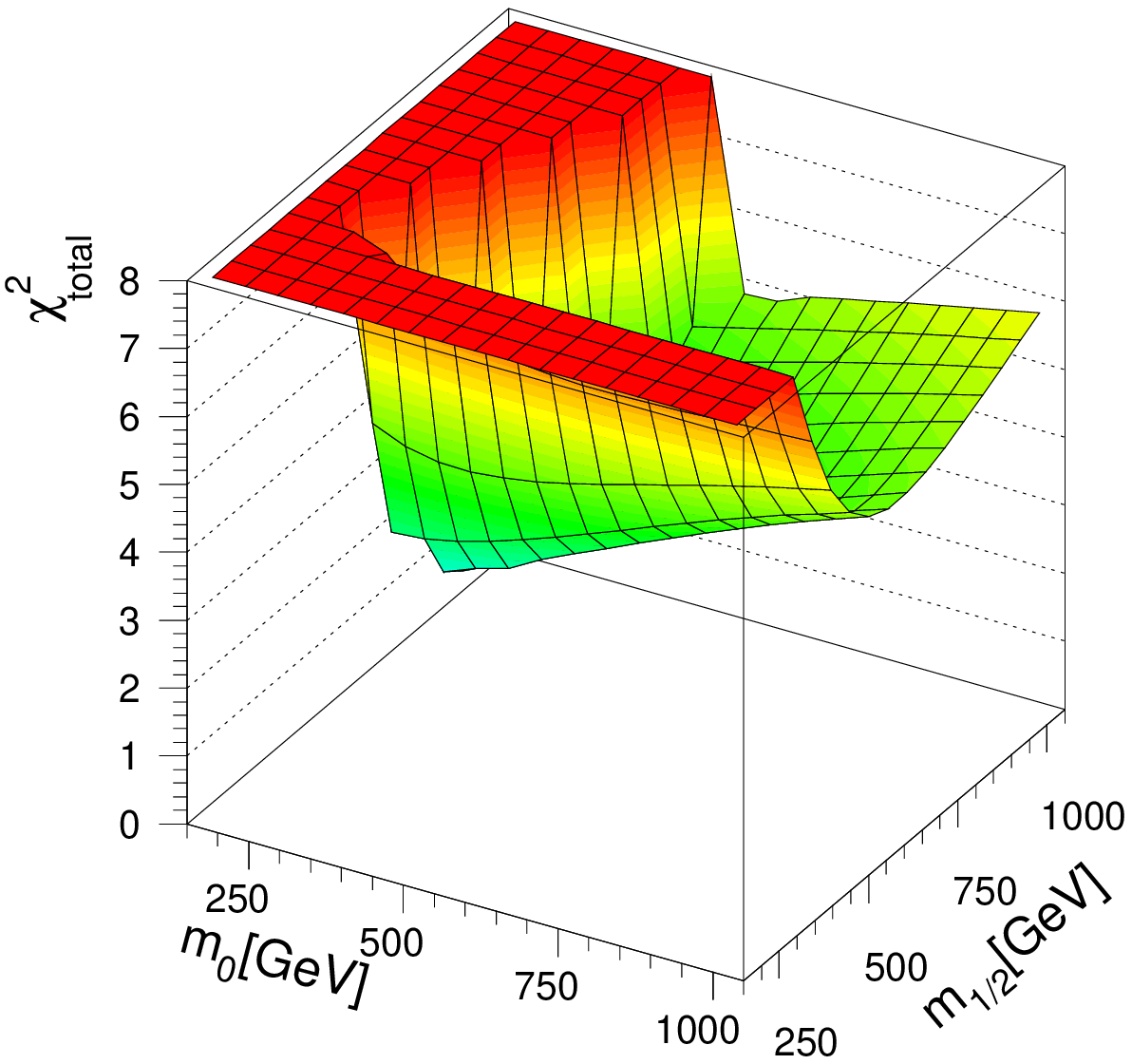}
    \includegraphics[width=0.4\textwidth]{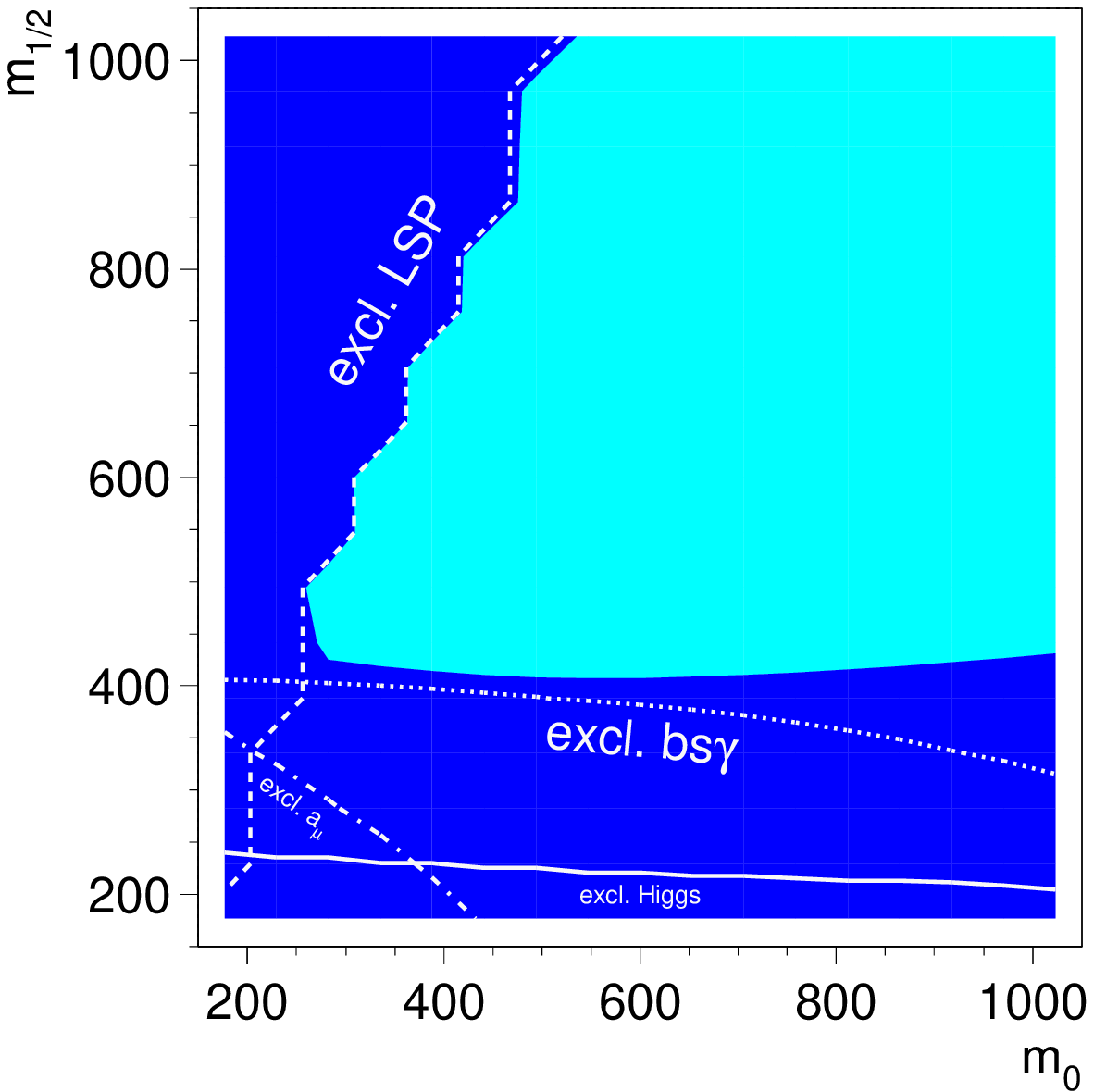}
    \caption[]{The total $\chi^2$ in the $m_0$-$m_{1/2}$-plane of a global fit performed in the CMSSM for different $\tan\beta=20,35,50$. Parameter regions with $\chi^2>\chi^2_{\mbox{\scriptsize{min}}}+2.7$ are excluded, which corresponds to a single sided 95\% confidence level. If
   $\Delta a_\mu^{small}$  instead of $\Delta a_\mu^{large}$ is used,
        the upper limits on $m_0$ and $m_{1/2}$ vanish as described in the text.}
      \label{chi2contour}
  \end{center}
\end{figure}

\begin{figure} [tbp]
  \begin{center}
    \includegraphics[width=0.32\textwidth]{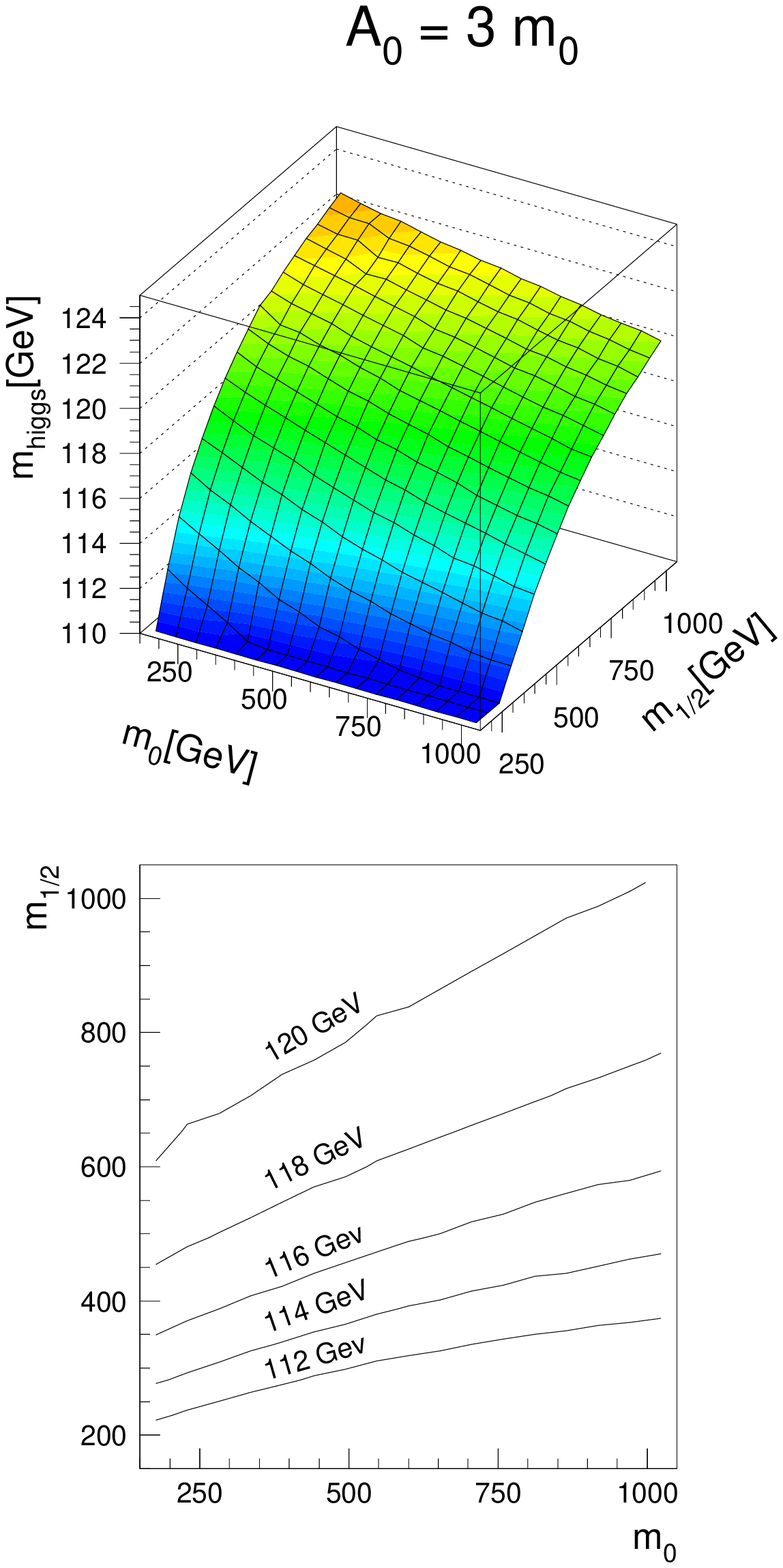}
    \includegraphics[width=0.32\textwidth]{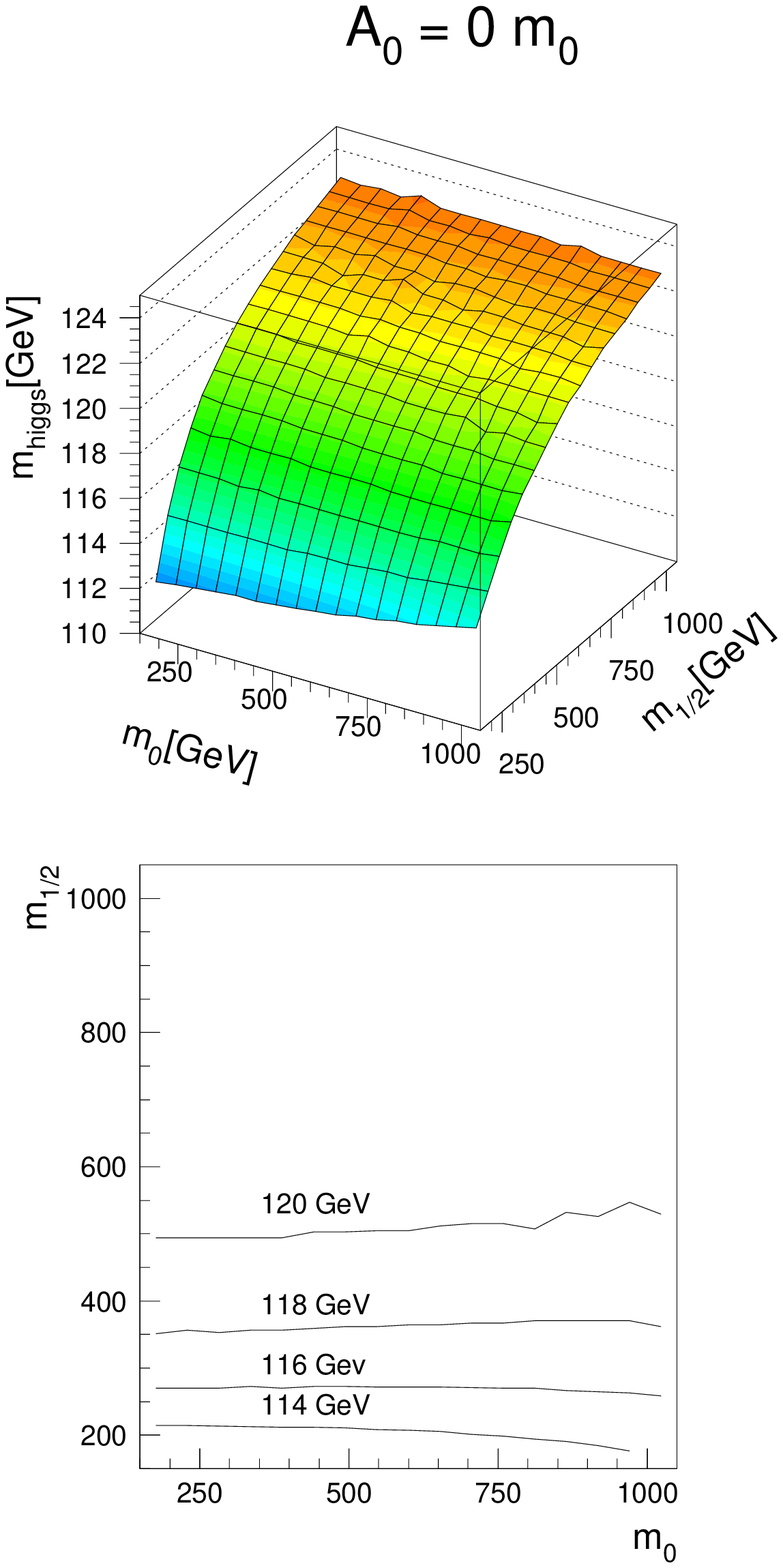}
    \includegraphics[width=0.32\textwidth]{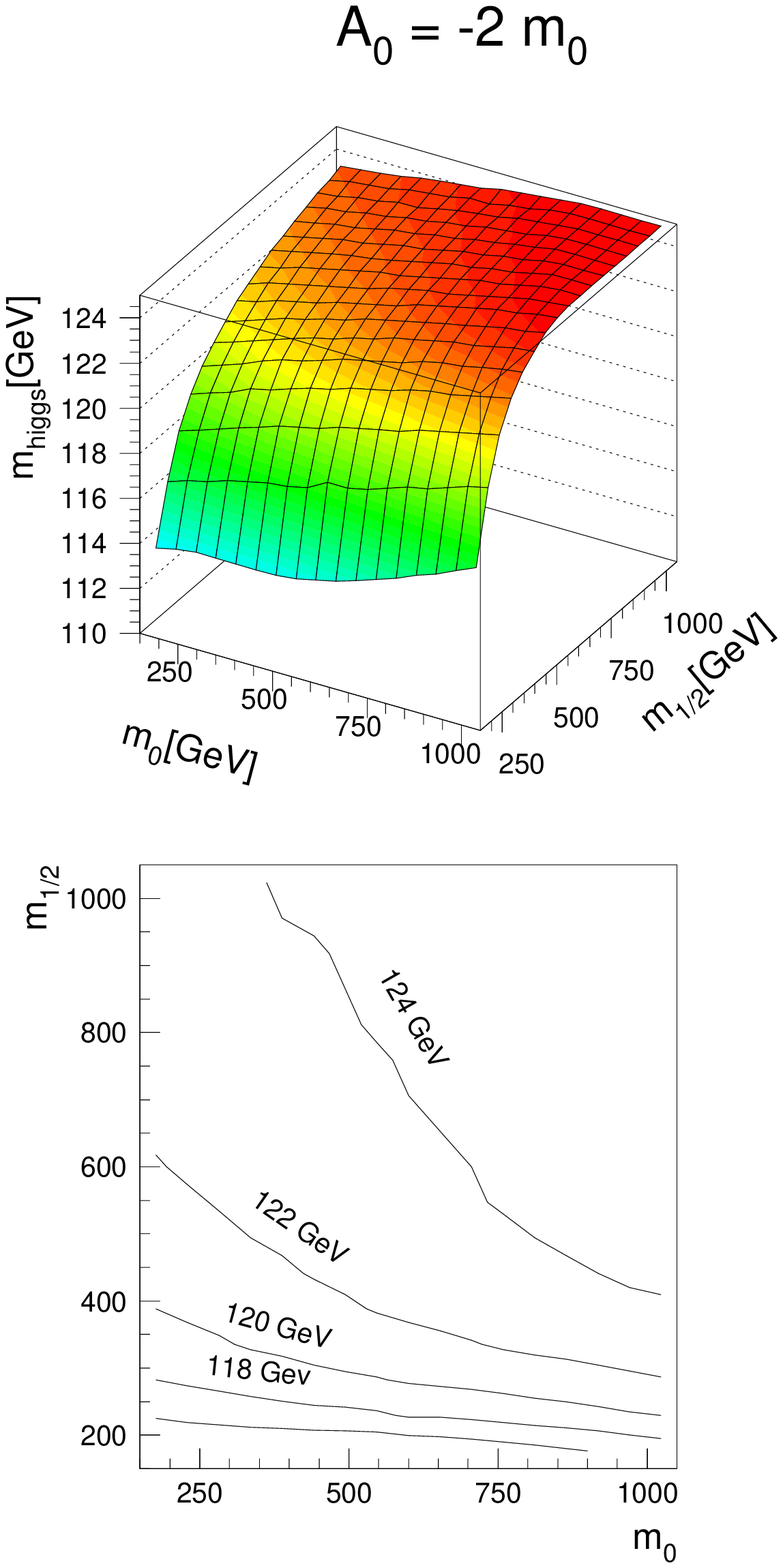}
    \caption[]{Dependence of Higgs mass on trilinear coupling $A_0$ in the $m_0$-$m_{1/2}$-plane. To calculate the Higgs mass the program package \texttt{FeynHiggsFast 1.2} was used \cite{FHF}.} \label{mh_aaa}
  \end{center}
\end{figure}

  In Fig. \ref{chi2contour} the allowed CMSSM parameter region in the $m_0$-$m_{1/2}$-plane is shown. One can see that a rather large region is allowed (light shaded region on the right hand side). In the fit the following constraints are included: the lightest supersymmetric particle (LSP) should be neutral, present Higgs limit ($m_h>114.6$ GeV), \bsg~ and $a_\mu$ \cite{bsgamu}. The trilinear coupling $A_0$ at the GUT scale is a free parameter in our fits with $\tan\beta=20$ and $\tan\beta=35$. The fit prefers positive values of $A_0$, in which case the Higgs limit becomes important as shown in Fig. \ref{mh_aaa}. Note that for $A_0 <0$ the Higgs limit becomes very
  weak, which is often not taken into account in the literature.
  If $A_0$ is fixed to zero\footnote{$A_0$ is the starting value at the GUT scale, so the low energy values are different.}, as in our fit with $\tan\beta=50$, the Higgs limit becomes less important but in exchange \bsg~ becomes the dominant lower limit. 

If the value of $\tan\beta$ becomes large, the mixing becomes large, as one can see in the off-diagonal elements in the $\tilde\tau$ mixing matrix:
\begin{equation}  
  \mathcal{M}^{\tilde \tau}=\left( 
    \begin{array}{cc}
      \tilde m^2_{\tau_L} & m_\tau(A_\tau-\mu\tan \beta) \\
      m_\tau(A_\tau-\mu \tan \beta) & \tilde m^2_{\tau_R}
    \end{array} \right)
    \label{mstau}
\end{equation}
In this case the lightest stau can easily get a mass below the lightest neutralino, especially
in regions where $m_{1/2} >m_0$, i.e. when the neutralinos before mixing  are already heavy
 compared with the stau.
Since the LSP is stable in the CMSSM, it should be neutral, so the stau cannot be the LSP
and these regions in the $m_{0}$-$m_{1/2}$ plane are excluded and shown is ``excl. LSP''
in Fig. \ref{chi2contour}.
%
%After running the RGE code, $\mu$ turns out to be larger if $\vert A_0\vert$ was large ($\vert A_0\vert\approx3\cdot m_0$), so we fixed $A_0$ to 0 for the fit with $\tan\beta=50$.
From the allowed parameter region in the $m_0$-$m_{1/2}$ plane, shown in Fig. \ref{chi2contour}, one can get limits on the neutralino and chargino masses, which are summarized in Tab. \ref{gauginolimit}.

\begin{table} [tbp]
  \begin{center}
    \begin{tabular}{|c|c|c|c|c|c|c|}
      \hline
      & \multicolumn{3}{|c|}{Neutralinos} & \multicolumn{3}{|c|}{Charginos} \\
      \hline
      $\tan\beta$ & lower & upper & exp. lower & lower & upper & exp. lower \\
      \hline
      6.5 &  93 & 134 & & 174 & 256 &\\
      20  & 123 & 292 & & 225 & 552 &\\
      35  & 135 & 382 & $\sim$50 (LEP2) & 248 & 690 & 103.5 (LEP2,$M_{\tilde{\nu}}< M_{{\tilde{\chi}}^\pm}$)\\
      50  & 169 & - & & 318 & - & \\
      \hline
    \end{tabular}
    \caption[]{Lower and upper limits on the Neutralino and Chargino masses using $Br(\bsg)$, $a_\mu$ and the light Higgs mass as constraints. The upper mass limits are corresponding to $\Delta a_\mu^{\mbox{\scriptsize{large}}}$. If one uses $\Delta a_\mu^{\mbox{\scriptsize{small}}}$ the upper limits will vanish. The present experimental lower mass limits are below the limits of our analysis, so new experimental data is awaited.} \label{gauginolimit}
  \end{center}
\end{table}

\clearpage

\section{Gauge unification and the strong coupling constant}

  In this section we reconsider the determination of the coupling constants from the electroweak fit and compare it with the coupling constants needed for unification. The gauge couplings in the $\overline{\mbox{MS}}$ scheme determining unification can be written as:

  \begin{eqnarray}
    \alpha_1 &=& (5/3) \alpha^{\overline{\mbox{\scriptsize{MS}}}}/\cos^2 \theta^{\overline{\mbox{\scriptsize{MS}}}}_W, \nonumber \\
    \alpha_2 &=& \alpha^{\overline{\mbox{\scriptsize{MS}}}} / \sin \theta^{\overline{\mbox{\scriptsize{MS}}}}_W, \nonumber\\
    \alpha_3 &=& \alpha_s^{\overline{\mbox{\scriptsize{MS}}}}, \nonumber
  \end{eqnarray}

\begin{figure} [tbp]
  \begin{center}
    \includegraphics[width=0.6\textwidth]{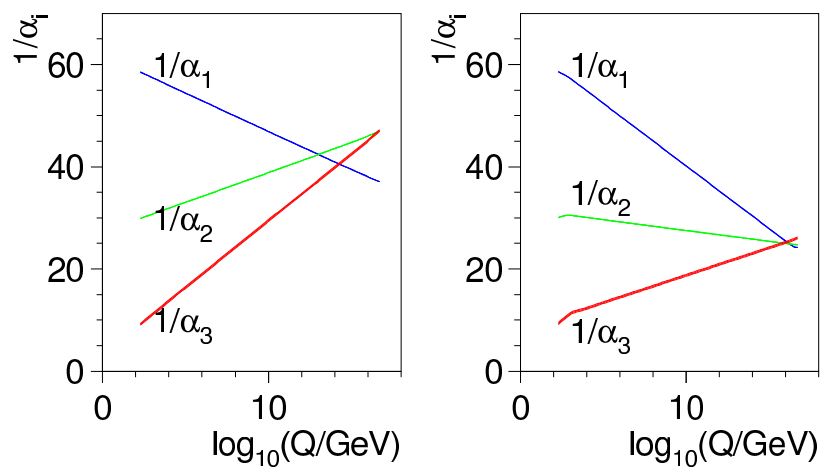}
    \caption[]{Here the running of the couplings in the SM (left) and MSSM (right) is shown. In the MSSM unification is possible due to threshold corrections of supersymmetric particles.} \label{unification}
  \end{center}
\end{figure}

\begin{figure} [tbp]
  \begin{center}
    \includegraphics[width=0.49\textwidth]{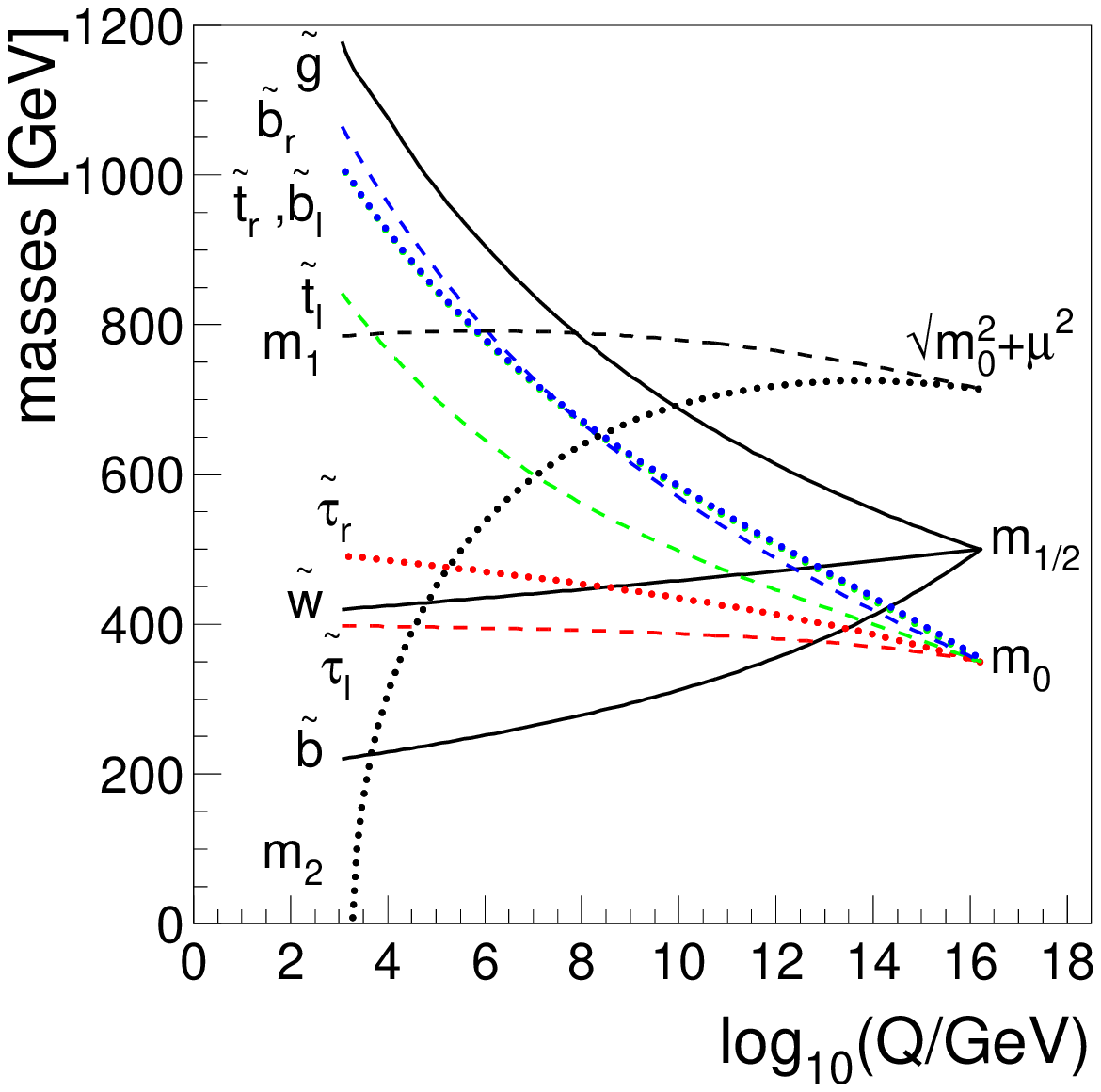}
    \includegraphics[width=0.49\textwidth]{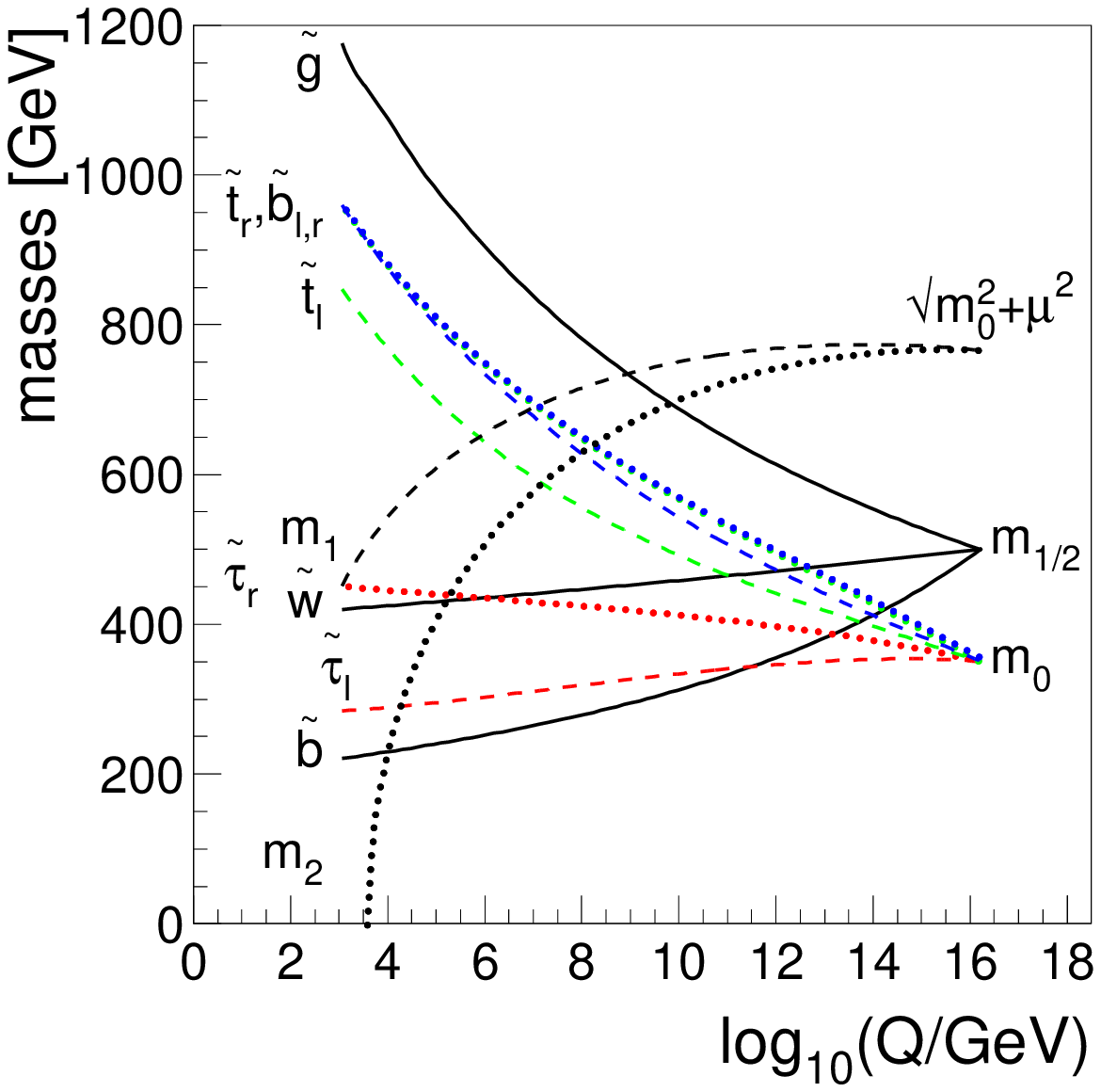}
    \caption[]{The evolution of the mass parameters  for the mSUGRA parameters $m_0=350$~GeV, $m_{1/2}=500$~GeV. In the left panel $\tan\beta=5$ and in the right one $\tan\beta=50$.  At the GUT scale gauginos are unified to $m_{1/2}$ and sfermions  to $m_0$, while the mass paramters in the Higgs potential start at $\sqrt{\mu^2+m_0^2}$ } \label{unification_masses}
  \end{center}
\end{figure}

\begin{figure} [tbp]
  \begin{center}
    \includegraphics[width=0.6\textwidth]{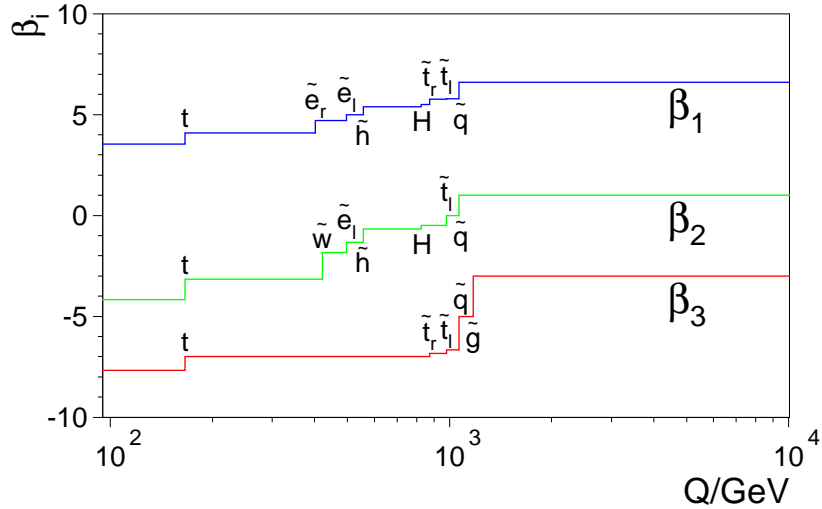}
    \caption[]{The energy dependence of the beta coefficients fo the first order
      Renormalization Group Equations for the running of  the gauge couplings.
      They different steps at the SUSY mass thresholds correspond to changes
      of the slopes in the running of the couplings in Fig. \ref{unification}.
      The mSUGRA parameters are choosen to be: $m_0=350$~GeV, $m_{1/2}=500$~GeV and $\tan\beta=50$.} \label{betas}
  \end{center}
\end{figure}

\begin{figure} [tbp]
  \begin{center}
    \includegraphics[width=0.49\textwidth]{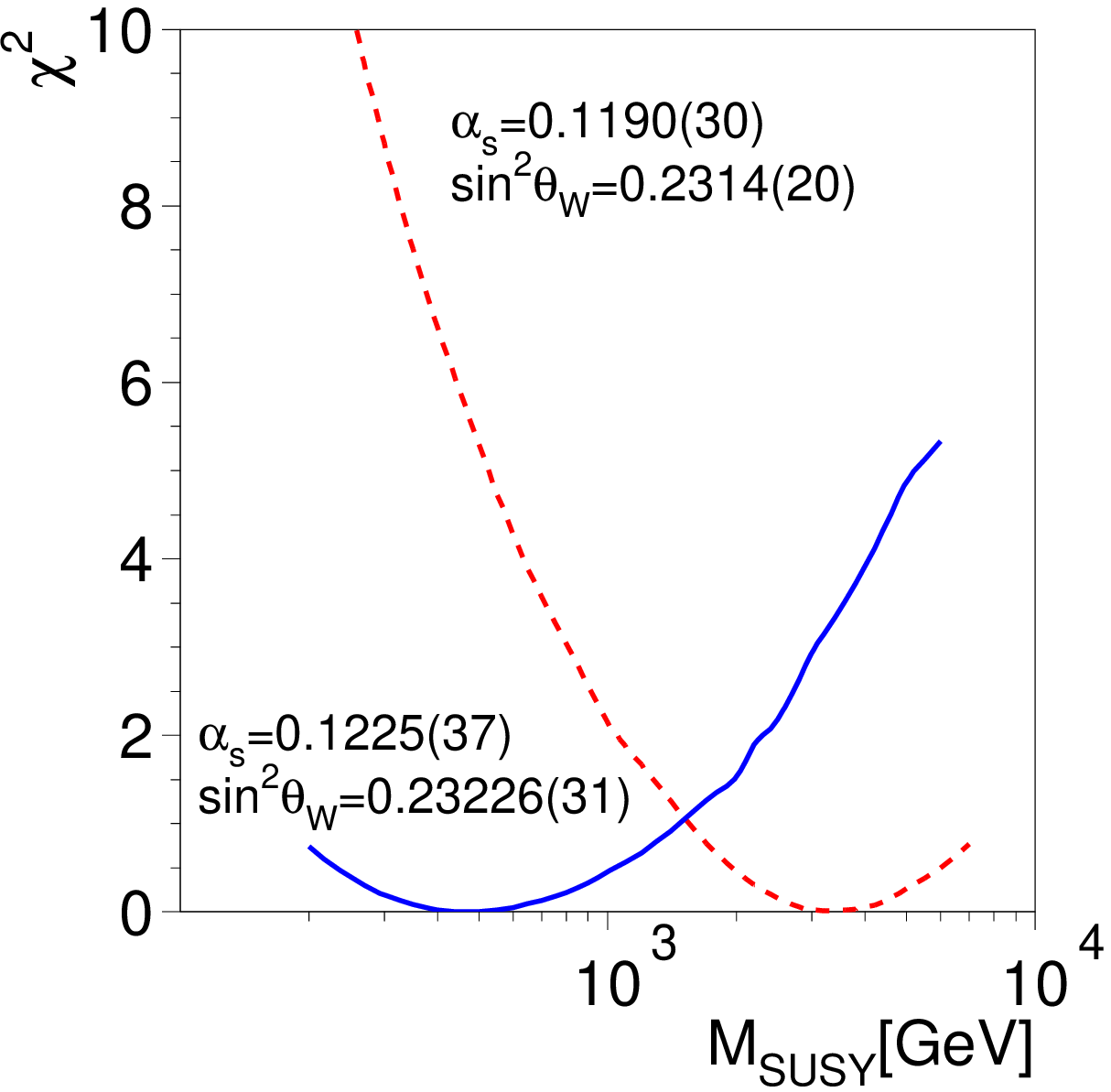}
    \includegraphics[width=0.49\textwidth]{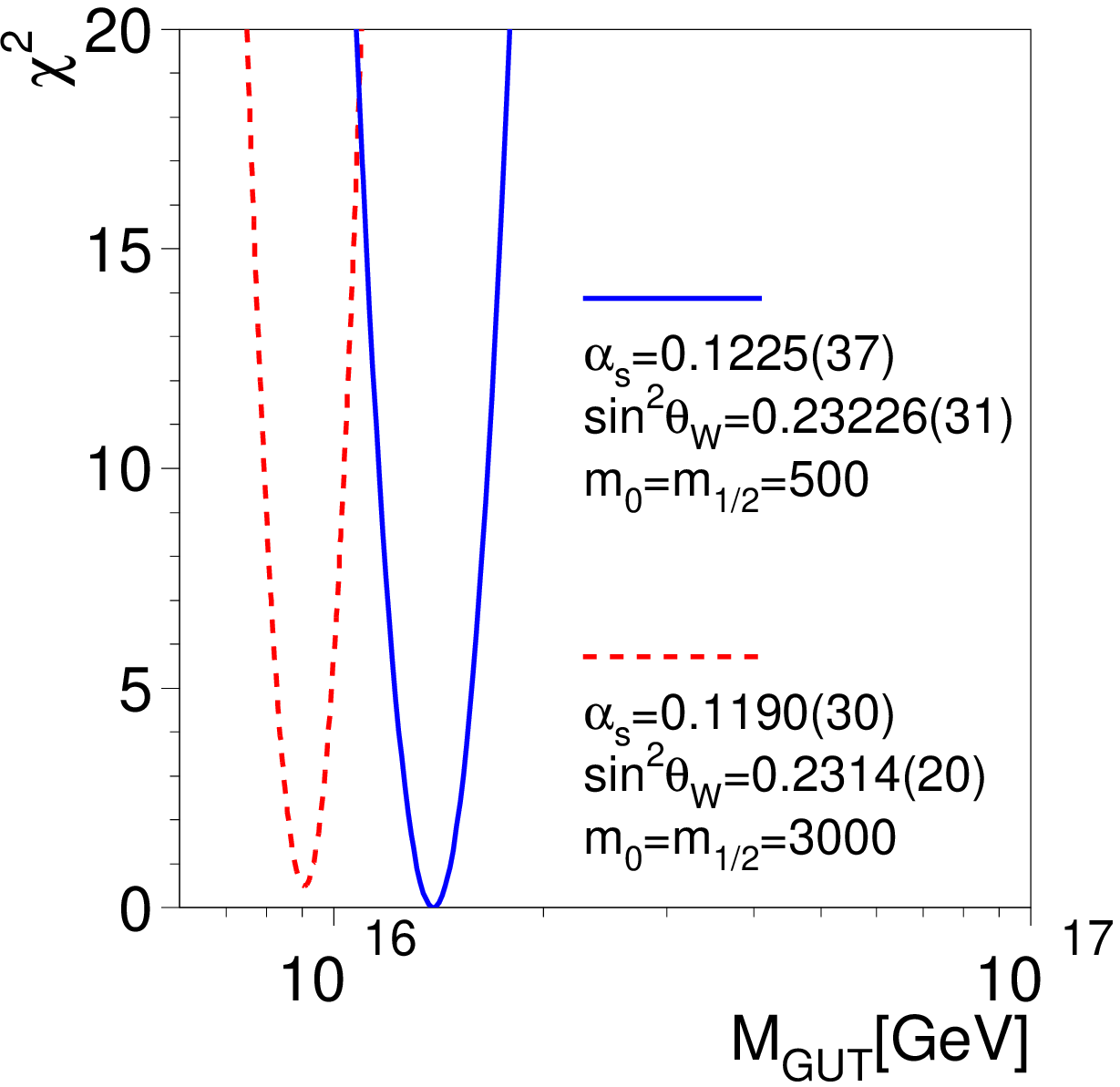}
    \caption[]{The dependence of $\chi^2_{\mbox{\scriptsize{unification}}}$ on $M_{\mbox{\scriptsize{SUSY}}}=m_0=m_{1/2}$ (left hand side) and on $M_{\mbox{\scriptsize{GUT}}}$ (right hand side) is shown. The two different sets of $\alpha_s(M_Z)$ and \sinw,
      which are discussed in the text, yield quite different SUSY masses needed for
      unification      ($\sim 500$~GeV to $\sim 3000$~GeV).} \label{chi2unification}
  \end{center}
\end{figure}

  In the MSSM gauge unification can be reached in contrast to the SM (see Fig. \ref{unification}). Instead of a common SUSY mass scale we use a more sophisticated mass spectrum \cite{carena}-\cite{langacker2}. The high energy mSUGRA parameters determine the low energy masses and couplings via RGEs. The running of the masses is shown in Fig. \ref{unification_masses} for low and high values of $\tan\beta$. The supersymmetric particles contribute to the running of the gauge couplings at energies above their masses as shown in Fig. \ref{betas}. The mass scale of SUSY particles and the unification scale \Mgut, which yields perfect unification is dependent on the low energy values of the gauge couplings (see Fig. \ref{chi2unification}).

  How good the gauge couplings can be unified at high energies depends on the experimental low energy values of them. We use the fine structure constant $\alpha(M_Z)=1/127.953(49)$ \cite{burkhardt}. The other ingredients at $M_Z$, the electroweak mixing angle \sinw~ and the strong coupling constant $\alpha_s$, are best determined from the electroweak precision data of the $M_Z$ line shape at LEP and SLC. Unfortunately the \sinw~ data disagree by about $3~\sigma$. Clearly, the SLC value yields a Higgs mass, which is below the present Higgs limit of 114.6 GeV, but the average value is consistent with it (see Fig. \ref{sw2mh}).

  In addition, the strong coupling constant depends on the observables used in the fit: if only $M_Z$, $\Gamma_{\mbox{\scriptsize{tot}}}$ and $\sigma_{\mbox{\scriptsize{had}}}^0$ are used, a value of $\alpha_s=0.115(4)$ is found as shown in Tab. \ref{alpha_s}, while the ratio $R_l$ of the hadronic and leptonic partial widths of the $Z_0$ boson yields a higher value $\alpha_s=0.123(4)$. 
%Note that this value is above the so-called world average quoted in the Partice Data Book of $\alpha_s=0.1185(27)$. However, it is important to realize that the QCD corrections to $R_l$ and $\sigma_{\mbox{\scriptsize{had}}}^0$ are calculated up to $\mathcal{O}(\alpha_s^3)$, whereas most of the variables used in the world average have not been calculated to this accuracy. Even for such an inclusive measurement as the total cross section, the third order QCD corrections are 2\% of the first order corrections and therefore non-negligible. 
Another quantity, which has been calculated up to $\mathcal{O}(\alpha_s^3)$ is the ratio of hadronic and leptonic widths of the $\tau$ lepton, $R_\tau$, which yields  a value close to the value from $R_l$: $\alpha_s=0.121(3)$.

\begin{table} [tbp]
  \begin{center}
    \begin{tabular}{|c|c|c|c|}
      \hline
      & all data & $M_Z$, $\Gamma_{\mbox{\scriptsize{tot}}}$, $\sigma_{\mbox{\scriptsize{had}}}^0$ & $R_l$ \\
      \hline
      $\alpha_s(M_Z)$   &0.1183(26)                &0.1154(40)         & 0.1225(37)\\
      $\sin^2\theta_W^{\overline{\mbox{\scriptsize{MS}}}}$
                        &0.23136(16)               &0.23148(2)         & 0.23152(2)\\
      $\Delta\alpha^{(5)}_{\mbox{\scriptsize{had}}}$
                        & 0.02770(31)              & 0.02761(36)(fixed)& 0.02761(36)(fixed)\\
      $\alpha(M_Z)^{-1}(\overline{\mbox{\small{MS}}})$
                        &127.931(42)               &127.953(49) (fixed)& 127.953(49)(fixed)\\
      $M_Z$             &91.1874(20) GeV           &91.1875 GeV(fixed) &91.1875 GeV(fixed)\\
      $M_H$             &$88^{+54}_{-35}$ GeV      &114.6 GeV(fixed)   &114.6 GeV(fixed)\\
      $m_t$             &175.5(4.1) GeV            &174.3 GeV(fixed)   &174.3 GeV(fixed)\\
      \hline
    \end{tabular}
  \end{center}
  \caption[]{Some SM parameter sets for different input data. If only $M_Z$, $\Gamma_{\mbox{\scriptsize{tot}}}$ and $\sigma_{\mbox{\scriptsize{had}}}$ are used $\alpha_s(M_Z)$ goes down to the lower limit of the present world average. If the parameters are determined from $R_l$, $\alpha_s(M_Z)$ goes to the upper limit.} \label{alpha_s}
\end{table}

\begin{table} [tbp]
  \begin{center}
    \begin{tabular}{|c|c|c|}
      \hline
      $\sigma_{\mbox{\scriptsize{had}}}^0$ & $\alpha_s$ & $N_\nu$ \\
      \hline
      $\pm 0 ~\sigma$ & 0.1154(40) & 2.982(8)\\
      $-1 ~\sigma$ & 0.1168(40) & 2.988(8)\\
      $-2 ~\sigma$ & 0.1182(40) & 2.995(8)\\
      $-3 ~\sigma$ & 0.1196(40) & 3.002(8)\\
      \hline
    \end{tabular}
  \end{center}
  \caption[]{The luminosity influences the hadronic cross section $\sigma_{\mbox{\scriptsize{had}}}^0$. If the luminosity is larger than measured than value of $\sigma_{\mbox{\scriptsize{had}}}^0$ decreases. This causes a increase in $\alpha_s$ and the number of neutrino families $N_\nu$. Note that a 2~$\sigma$ deviation in $N_\nu$ corresponds to a 3~$\sigma$ deviation in $\sigma_{\mbox{\scriptsize{had}}}^0$.} \label{change_sig}
\end{table}

\begin{figure} [tbp]
  \begin{center}
    \includegraphics[width=0.3\textwidth]{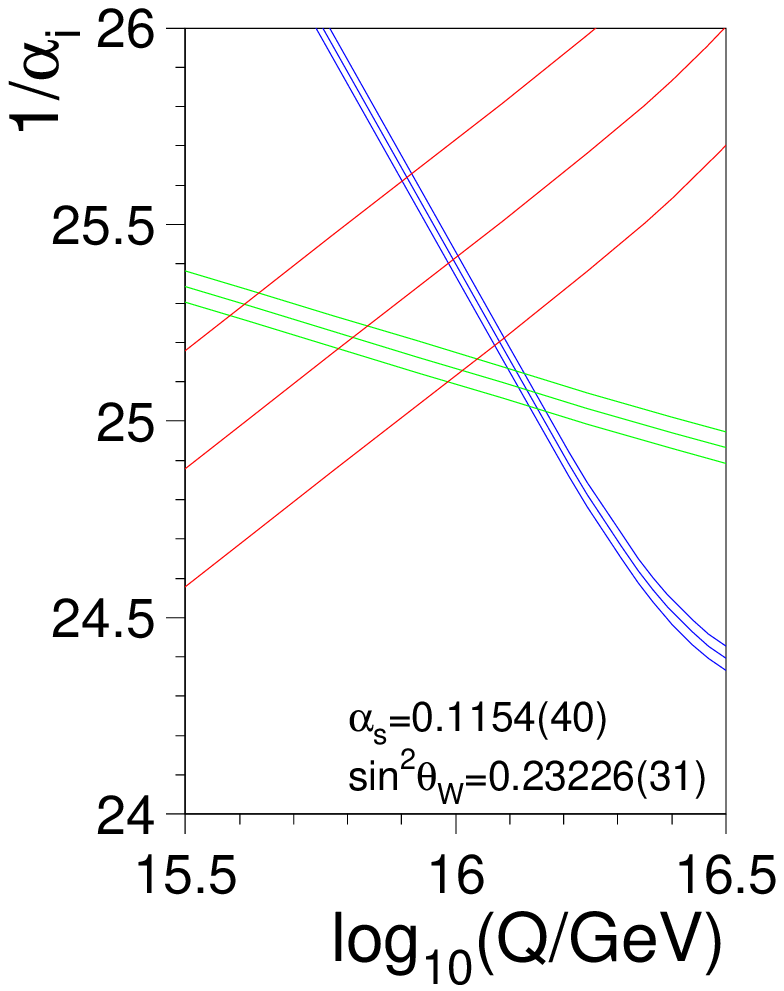}
    \includegraphics[width=0.3\textwidth]{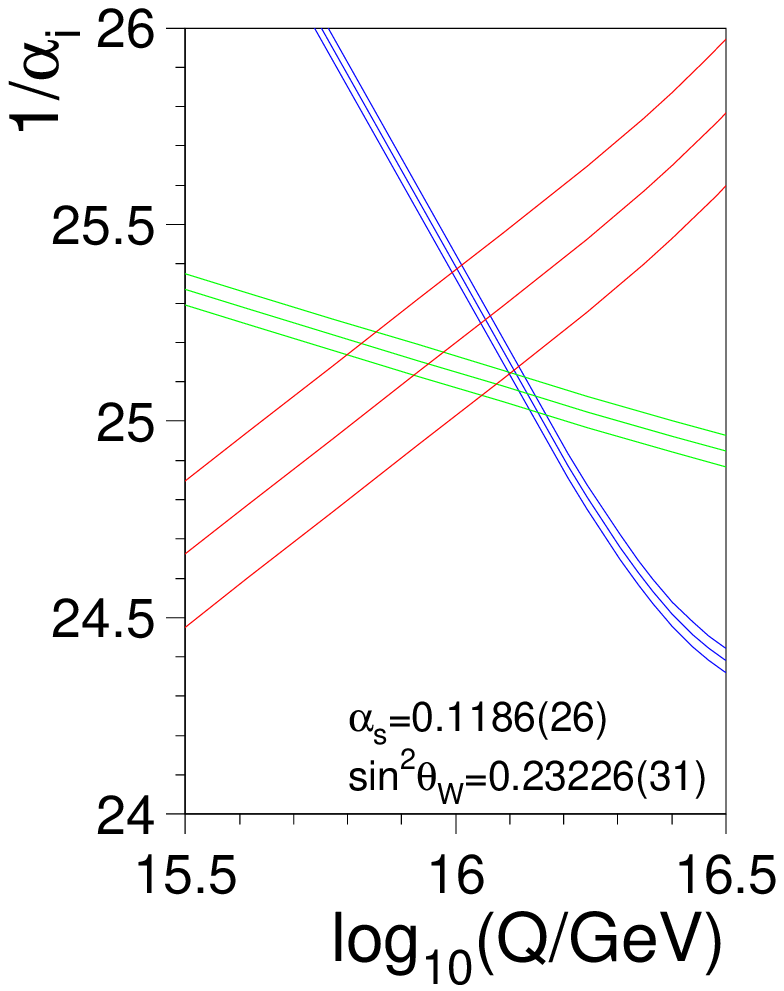}
    \includegraphics[width=0.3\textwidth]{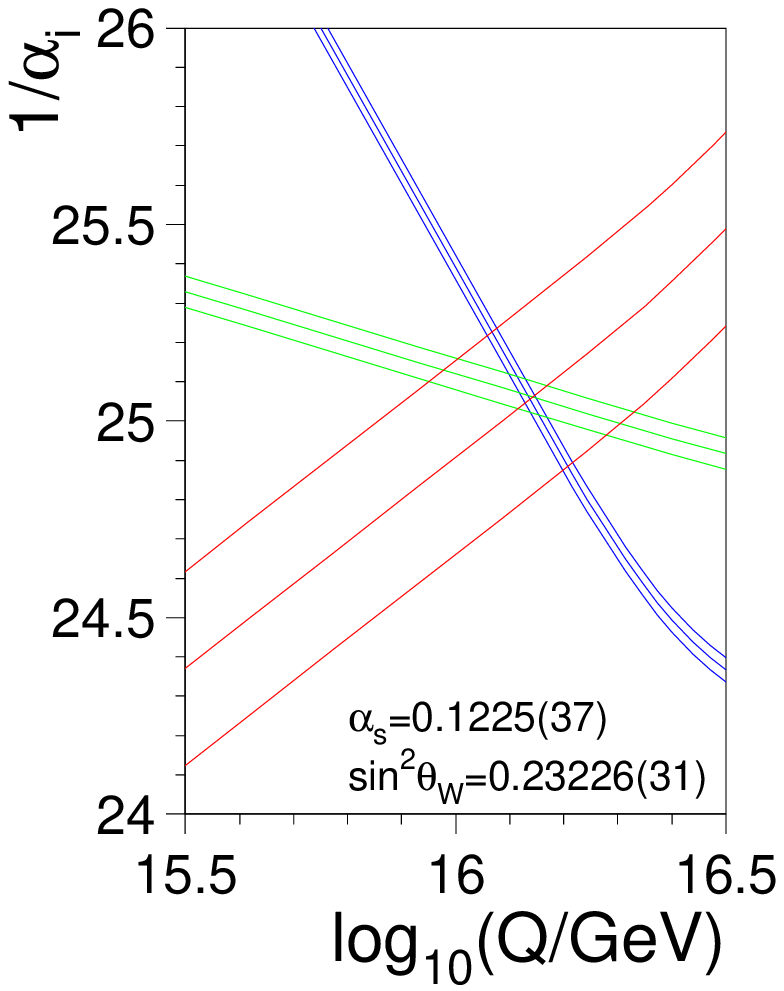}
    \includegraphics[width=0.3\textwidth]{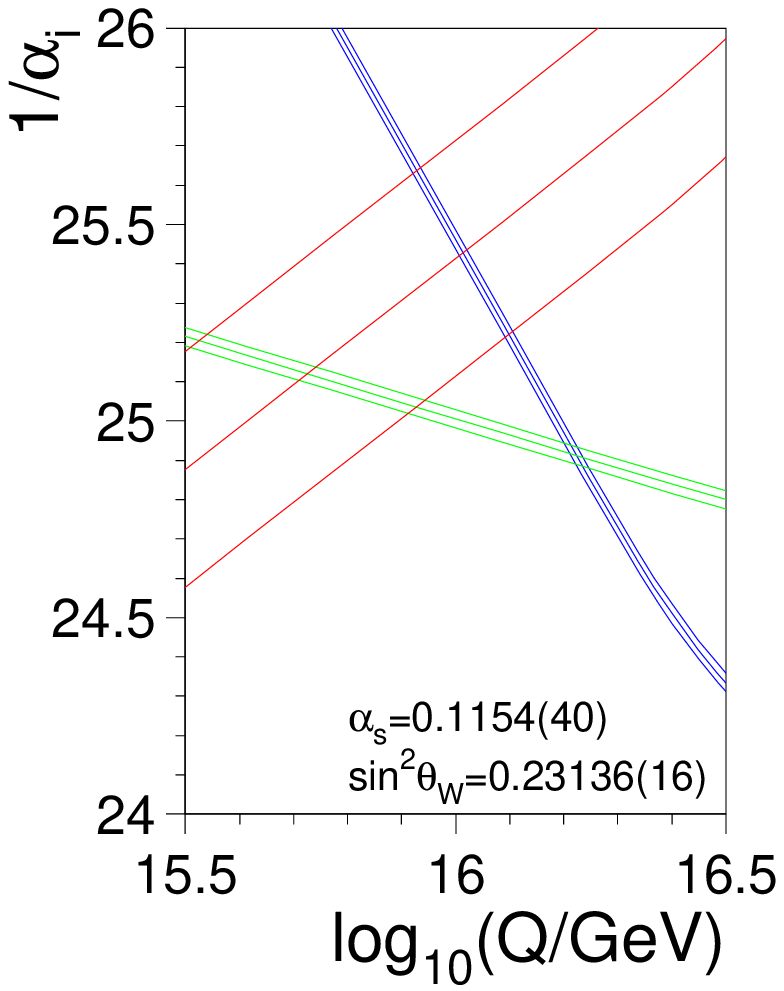}
    \includegraphics[width=0.3\textwidth]{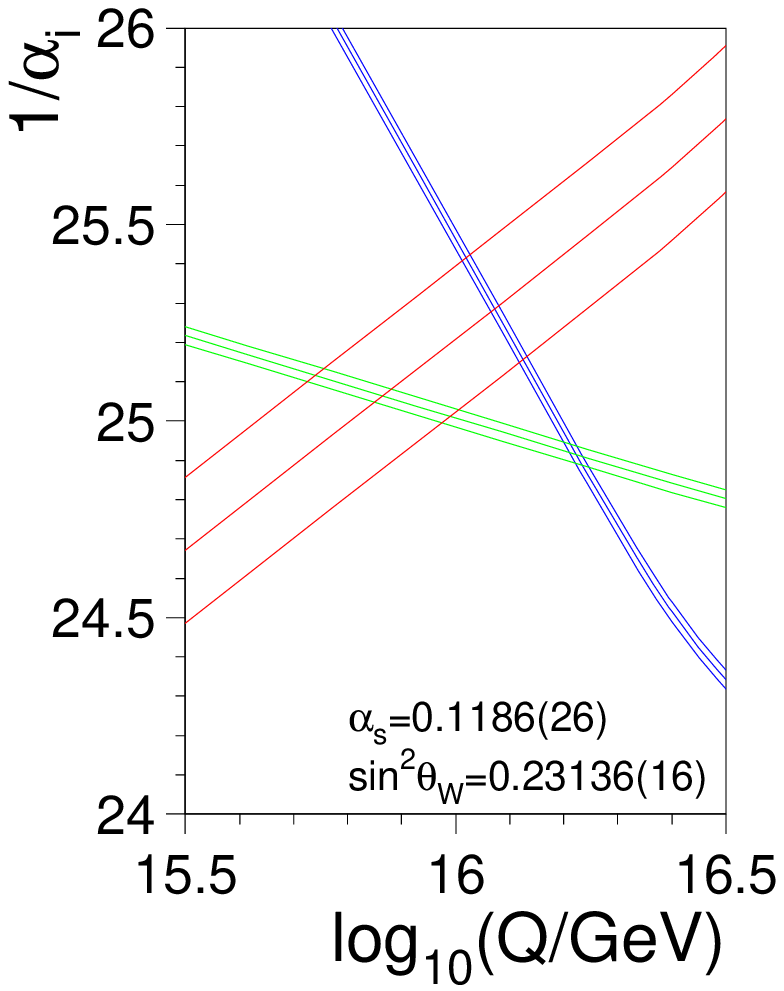}
    \includegraphics[width=0.3\textwidth]{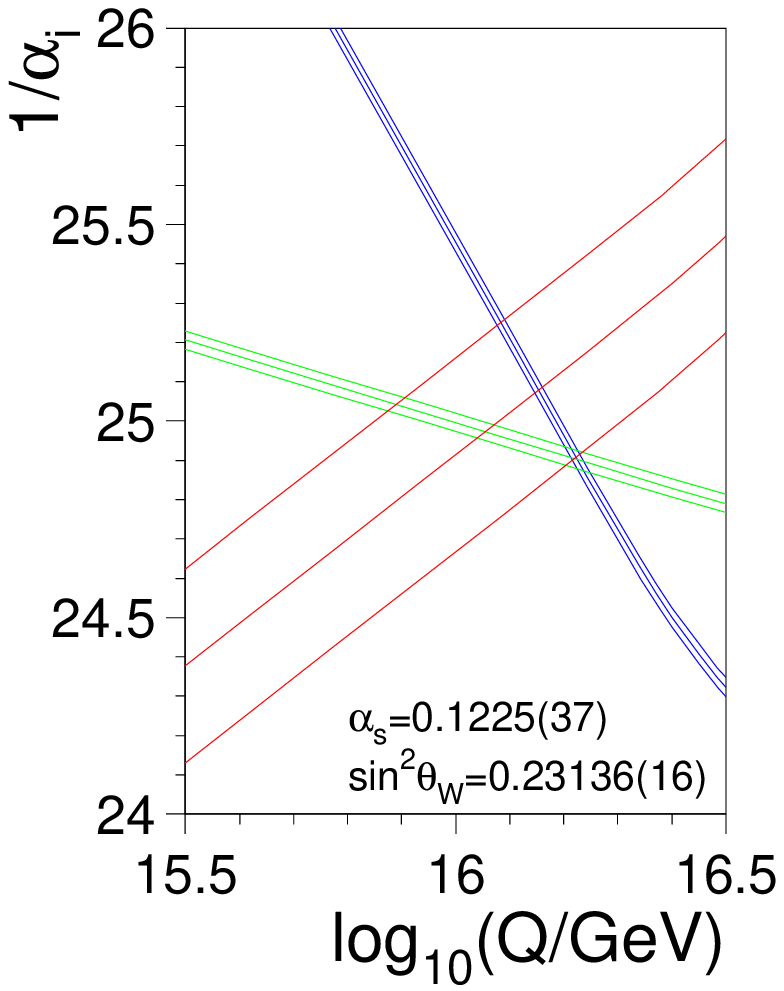}
    \includegraphics[width=0.3\textwidth]{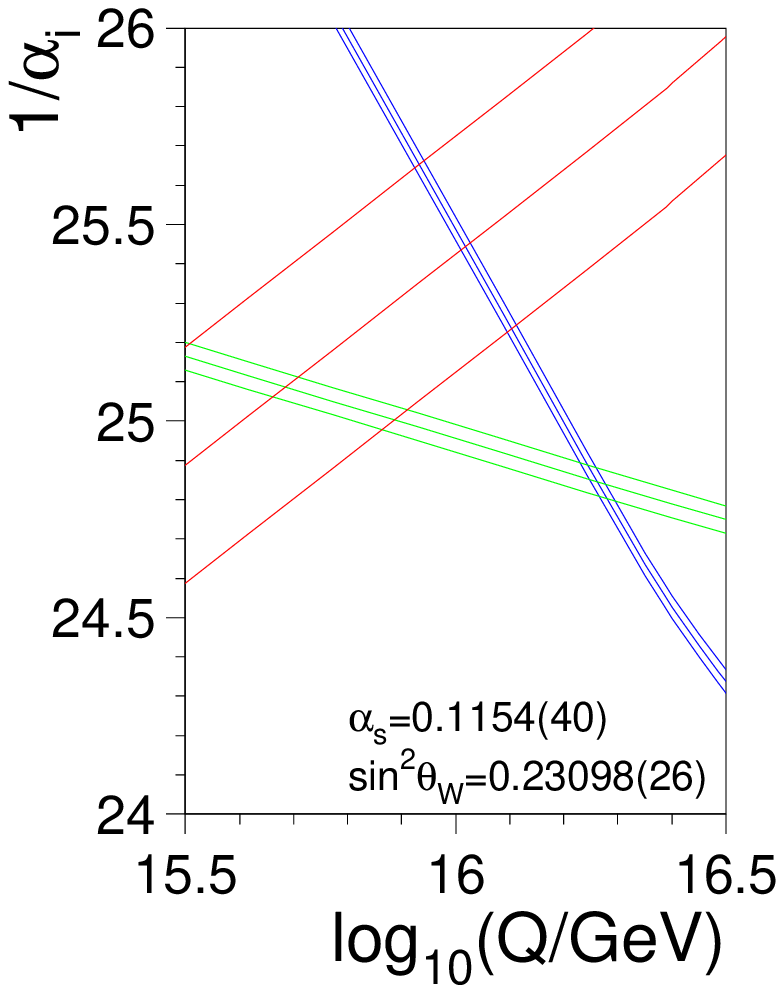}
    \includegraphics[width=0.3\textwidth]{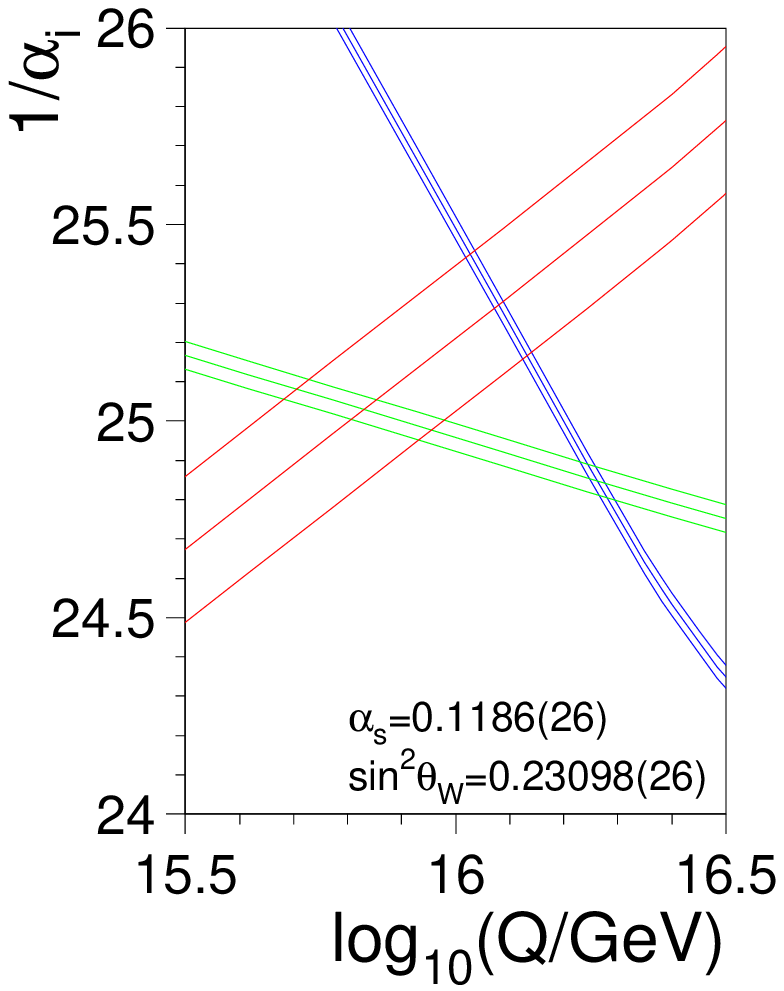}
    \includegraphics[width=0.3\textwidth]{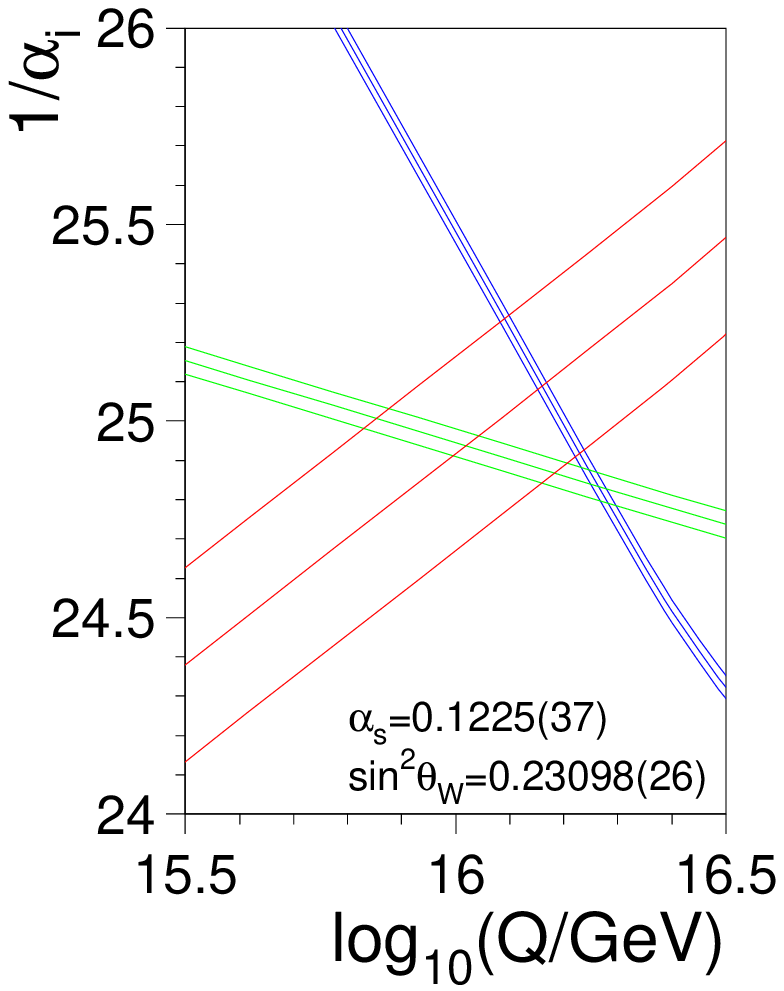}
    \caption[]{Zoom of the region, where the gauge couplings unify in the MSSM. For the plots different $\alpha_s(M_Z)$ and \sinw~ are used. In the left column we use $\alpha_s(M_Z)$ from $\sigma_{\mbox{\scriptsize{had}}}^0$, the middle one the world average and in the right one from $R_l$. In the first row we use \sinw~ from $A_{FB}^b$, in the middle one the world average and in the lowest one from $A_l$. In these fits the mSUGRA parameters $m_0=350$~GeV, $m_{1/2}=500$~GeV and $\tan\beta=50$ are used.} \label{unification_fine}
  \end{center}
\end{figure}
\begin{figure} [tbp]
  \begin{center}
    \includegraphics[width=0.5\textwidth]{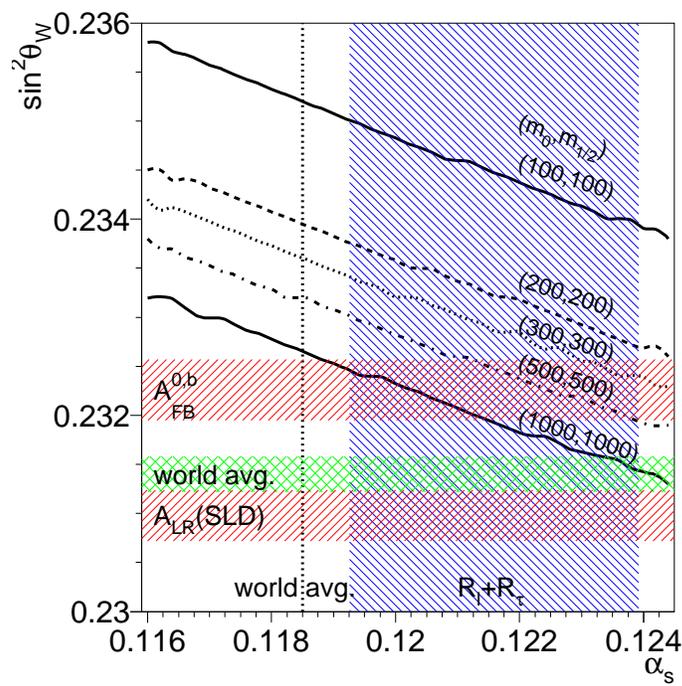}
    \caption[]{The lines are combinations of $\alpha_s(M_Z)$ and \sinw, which
      yield gauge unification. Unification with the world average for \sinw~
      requires $\alpha_s>0.123$, if SUSY mass scales are required to be below
      1 TeV. The value is close to $\alpha_s=0.122(2)$, which is the average
      from $R_l$ and $R_\tau$, the only measured ratios calculated up to
      order $\mathcal{O}(\alpha_s^3)$ in QCD. Note that this value is high compared with the measurement from the $Z_0$ hadronic peak cross section $\alpha_s=0.115(4)$, but the latter is dominated by the common luminosity errors from LEP. The too low number of neutrino species, as measured at LEP, suggests that the luminosity is off by $3\sigma$. If true, perfect gauge unification can be obtained for SUSY mass scales below 1 TeV.} \label{as_sinw}
  \end{center}
\end{figure}

\newpage
  The low  $\alpha_s$  value from the hadronic cross section may 
be due 
to an underestimated error of the luminosity as suggested by the fact that the electroweak fit yields a too low value of the number of neutrinos. This can be understood as follows: The line shape can be described by the following independent parameters \cite{ewwgreport}:
\begin{eqnarray}
  \sigma^0_{\mbox{\scriptsize{had}}}&=&{12\pi\over M_Z^2}{\Gamma_{ll}\Gamma_{\mbox{\scriptsize{had}}}\over\Gamma_Z^2} \nonumber \\
  R_l&=&{\Gamma_{\mbox{\scriptsize{had}}}\over\Gamma_{ll}} \nonumber \\
  \Gamma_Z&=&\Gamma_{\mbox{\scriptsize{had}}}+3\cdot\Gamma_{ll}+\Gamma_{\mbox{\scriptsize{inv}}} \nonumber
\end{eqnarray}

Instead of $\sigma^0_{\mbox{\scriptsize{had}}}$, $R_l$ and $\Gamma_Z$ one can use $\Gamma_{ll}$, $\Gamma_{\mbox{\scriptsize{had}}}$ and $\Gamma_{\mbox{\scriptsize{inv}}}$ as lineshape parameters. The transformation is:

\begin{eqnarray}
   \Gamma_{ll}&=& \sqrt{{M_Z^2\over 12\pi}{\sigma^0_{\mbox{\scriptsize{had}}}\over R_l}}\cdot\Gamma_Z\nonumber \\
   \Gamma_{\mbox{\scriptsize{had}}}&=&  \sqrt{{M_Z^2\over 12\pi}\sigma^0_{\mbox{\scriptsize{had}}} R_l}\cdot\Gamma_Z\nonumber \\
   \Gamma_{\mbox{\scriptsize{inv}}}&=& \left(1-3\cdot\sqrt{{M_Z^2\over 12\pi}{\sigma^0_{\mbox{\scriptsize{had}}}\over R_l}}-\sqrt{{M_Z^2\over 12\pi}\sigma^0_{\mbox{\scriptsize{had}}} R_l} \right)\cdot\Gamma_Z \nonumber
\end{eqnarray}

  The decay of a $Z$ into two neutrinos is invisible. That means, that from the invisible decay width $\Gamma_{\mbox{\scriptsize{inv}}}$ one can get the number of the light neutrino flavours $N_\nu$, which is given by:

$$N_\nu={\Gamma_{\mbox{\scriptsize{inv}}}\over\Gamma_{ll}}\cdot\left({\Gamma_{ll}\over\Gamma_{\nu\nu}}\right)_{\mbox{\scriptsize{SM}}}=\left(\sqrt{12\pi\over M_Z^2 \sigma^0_{\mbox{\scriptsize{had}}} R_l}-3-R_l \right)\cdot\left({\Gamma_{ll}\over\Gamma_{\nu\nu}}\right)_{\mbox{\scriptsize{SM}}}$$

  The LEP I results yield $N_\nu=2.982(8)$, which is 2~$\sigma$ below the expected value of 3. The origin of this deviation could be $\sigma^0_{\mbox{\scriptsize{had}}}$ or $R_l$, since the SM ratio of leptonic and neutrino decay width is constant. The $\sigma^0_{\mbox{\scriptsize{had}}}$ measurements of all four LEP experiments aggree within 1~$\sigma$. So if there is a shift, than it is due to common systematic errors. One should note that the error of the observable $\sigma_{\mbox{\scriptsize{had}}}^0$, when averaged between all experiments, is dominated by the theoretical error of the luminosity, which is common for all experiments. All experiments use the same program code to calculate the luminosity via Bhabha scattering called BHLUMI \cite{bhlumi}. In contrast the ratio $R_l$ depends not on the luminosity.

If we assume that the luminosity, measured at LEP I, is too low, $\sigma^0_{\mbox{\scriptsize{had}}}$ will decrease if the luminosity is increased. A lower value of $\sigma^0_{\mbox{\scriptsize{had}}}$ yields a higher value of $\alpha_s$ and increases $N_\nu$. The numerical effects are exemplified in Table \ref{change_sig}: if the hadronic cross section is decreased by 3~$\sigma$, $\alpha_s$ is increased to a value close to the value obtained from $R_l$ and $R_\tau$, and $N_\nu$ becomes 3 in agreement with the fact that we know there are three different neutrinos.

In summary, the simultaneous problems of the too low number of neutrino
flavours  and the too low value of $\alpha_s$ from the hadronic cross section can be solved by the assumption of a common systematic error on the luminosity between all experiments. Increasing the hadronic cross section by $3~\sigma$ yields the correct number of neutrinos and solves the discrepancies between the $\alpha_s$ values from the cross section and $R_l$. Note that the latter observable is independent of the luminosity and yields $\alpha_s$ values in agreement with the hadronic tau decays. 

The unification of the gauge couplings for different pairs of $\alpha_s$ and \sinw~ is shown in Fig. \ref{unification_fine}. One can see, that perfect unification is possible for high values of $\alpha_s$ and \sinw.

The combination of \sinw~ values and $\alpha_s$ values needed for unification are indicated by the diagonal lines for different SUSY mass scales in Fig. \ref{as_sinw}. The averaged $\alpha_s$ values from $R_l$ and $R_\tau$ are indicated by the vertical band in Fig. \ref{as_sinw}. This band is above the world average, which is dominated by the hadronic peak cross section, but agrees better with the requirement of gauge unification in the CMSSM, as can be seen from Fig. \ref{as_sinw}: if in addition \sinw~ from $A_{FB}^b$ is used, perfect unification is easily possible, for GUT scale SUSY masses below 1000 GeV. For \sinw~ from $A_{LR}$ no unification is possible with the central value of $\alpha_s$. Since this value is inconsistent with the value of \sinw~ from $A_{FB}^b$ at the $3\sigma$ level, new data from a future linear collider are eagerly awaited.

%  Given the fact that $R_l$ and $R_\tau$ are ratios, they are not dependent on the errors of the luminosity in contrast to $\sigma_{\mbox{\scriptsize{had}}}^0$. Since $R_l$ and $R_\tau$ are the only two ratios, for which the QCD calculations have been calculated up to third order, we have indicated the average of these two measurements as the vertical band in Fig. \ref{as_sinw}. This band is clearly slightly above the world average, indicated by the vertical line, but agrees better with the requirement of unfication in the CMSSM.

  Clearly if \sinw~ from $A_{FB}^b$ and $\alpha_s$ from $R_l$ and $R_\tau$ are used, perfect unification is easily possible as shown in Fig. \ref{unification_fine} and Fig. \ref{as_sinw}. For the selection of the $\alpha_s$ values one has a good argument since there are the data for which $\mathcal{O}(\alpha_s^3)$ QCD calculations are available and since they are ratios, they are less sensitive to systematic errors. For \sinw~ there is no good argument to select parts of the data; on the other hand the data are inconsistent ($>3~\sigma$ apart). Here new data at a future linear collider are eagerly awaited.

%$$C=1+{\alpha_s\over\pi}+1.411\cdot\left({\alpha_s\over\pi}\right)^2-12.8\cdot\left({\alpha_s\over\pi}\right)^3+\cdots$$
%$$R_\tau=3.058\cdot\left[ 1+{\alpha_s(m_\tau)\over\pi}+5.2\cdot\left({\alpha_s(m_\tau)\over\pi}\right)^2+26.4\cdot\left({\alpha_s(m_\tau)\over\pi}\right)^3+\cdots\right]$$

\section{Conclusion}

  It has been shown that a SM electroweak fit including the anomalous magnetic moment of the muon $a_\mu$ and the branching ratio $Br(\bsg)$ yields a probability of about 5\%. The total $\chi^2$ is improved in the MSSM, mainly because of $a_\mu$, but the probability does increase only marginally due to the larger number of free parameters in the MSSM. However, in both cases the $3~\sigma$ discrepancy in \sinw~ from $A_{FB}^b$ and $A_{LR}$ is contributing to  the low probability. Since at present no arguments to doubt any of the measurements can be found, we tested the Particle Data Group's procedure to rescale the errors of these two measurements by the corresponding pulls. This yields considerably improved  $\chi^2$ values, both in the SM and MSSM, without significantly changing the fitted parameters.

  The electroweak data considerably constrain the CMSSM parameter space: the value of $\tb>6.5$
  and the gaugino mass parameter at the
  GUT scale has to be above $\approx 400$ GeV, which implies   a lower limit
  of 95 (175) GeV for neutralinos (charginos).
  
With the high precision values of the coupling constants  gauge unification
was reconsidered. The 3~$\sigma$ discrepancy in the values of the weak mixing
angle and the differences in $\alpha_s$  from the peak cross section of the
$Z^0$ resonance and  $R_l$, the ratio of hadronic and leptonic $Z_0$ decay
width, clearly effect unification. 
Unification prefers the larger values of $\alpha_s$ and \sinw. 
For the lowest value of \sinw~ from the
left-right asymmetries at the SLC ($A_{LR}$) no exact gauge unification is
possible. This low value of \sinw~  also corresponds to a
Higgs value of  40 GeV, which is below the lower limit of 114 GeV set 
by the direct searches.

The most reliable values of $\alpha_s$ are obtained from the ratios of hadronic and leptonic width of $Z^0$ and $\tau$-decays,  since they  do not depend on normalization uncertainties.
In contrast, $\alpha_s$  from the peak cross section  depends on the luminosity error.
Since all LEP experiments used the same luminosity Monte Carlo generator, the
final hadronic cross section  error is dominated by the common systematic uncertainty in the luminosity.
%The low value of $\alpha_s=0.115$ from the peak cross section and the  xxxx
% The error on $\sigma_{\mbox{\scriptsize{had}}}^0$ is dominated by the theoretical uncertainty on the luminosity, which is common to all experiments, since all experiments use the same luminosity MC generator.
  Lowering $\sigma_{\mbox{\scriptsize{had}}}^0$ by 3~$\sigma$ brings $\alpha_s$ from $0.115(4)$
   to 0.12 and $N_\nu$ from 2.982(8)  to 3, which is in better
agreement with the value $\alpha_s=0.123(4)$ from $R_l$ and the known 3
   neutrino flavours.
  For $\alpha_s > 0.12$ perfect unification can be  obtained much more easily. 
%ne finds perfect unification is possible for moderate SUSY masses ($<$ 1 TeV), especially for the larger electroweak mixing angle from LEP data. 
%Also for the averaged value of \sinw~ from these two measurements, for which one cannot find objective criteria to reject one or the other, exact gauge unification is possible within the errors.

\section{Acknowledgements} 

  The authors thank S. Jadach for interesting discussions about his luminosity
 program used by all LEP collaborations and possible systematic errors.
% We agree completely with him that it is a serious drawback to have only a single luminosity program with such high precision available for analysing the LEP data, since with the high LEP statistics the systematic luminosity error is the dominant error in the total hadronic cross section. 
\section{Note added in Proof}
The discrepancy between the   $\tau$ data 
and $e^+e^-$ hadronic cross sections, which change the SM prediction
of the anomalous magnetic moment (see section 2) seems to be settled
by preliminary KLOE data in favour of the $e^+e^-$ data, which are
in perfect agreement with CDM-2 data, as presented by Dr. Denig
at the open session of the LNF scientific committee 
(http://www.lnf.infn.it/committee/talks/26denig.pdf).
This strongly favours a supersymmetric contribution to the anomalous 
magnetic moment.

\end{document}